\newcommand{\hii}{H{\sc ii}}

\documentclass{aa}  
\usepackage{natbib}

\usepackage{xcolor}
\usepackage{float}
\usepackage{placeins}
\usepackage{graphicx}
\usepackage{gensymb}
\usepackage{multirow}
\usepackage{txfonts}
\usepackage{subfigure}
\usepackage{fixltx2e}
\usepackage[colorinlistoftodos]{todonotes}
\usepackage[flushleft]{threeparttable}
\usepackage[normalem]{ulem}
\usepackage{dcolumn}
\usepackage{csquotes}

\newcolumntype{d}[1]{D{.}{\cdot}{#1}}
\newcolumntype{.}{D{.}{.}{-1}}

\begin{document}

 
   \title{Studying the effects and cause of the massive star formation in Messier 8 East}
  
   \subtitle{}

   \author{M.\,Tiwari\inst{1}\fnmsep\inst{2}\fnmsep\thanks{Member of the International Max Planck Research School (IMPRS) for Astronomy and Astrophysics at the Universities of Bonn and Cologne.}\fnmsep, K.\,M.\,Menten\inst{1}, F.\,Wyrowski\inst{1}, A.\,Giannetti\inst{3},  M.-Y.\,Lee\inst{4}, W.\,-J.\,Kim\inst{5}, J.P.\,P\'{e}rez-Beaupuits\inst{6}} 
   

   \institute{Max-Planck Institute for Radio Astronomy,
              Auf dem H\"{u}gel, 53121 Bonn, Germany\\
              \email{mtiwari@mpifr-bonn.mpg.de}
              \and Department of Astronomy, University of Maryland, College Park, MD 20742-2421, USA\\
              \email{mtiwari@umd.edu}
              \and INAF-Istituto di Radioastronomia, and Italian ALMA Regional Centre, Via P. Gobetti 101, 40129 Bologna, Italy
              \and Korea Astronomy and Space Science Institute Daedeokdae-ro 776, Yuseong-gu Daejeon 34055, Republic of Korea
              \and Instituto de Radioastronom\'ia Milim\'etrica, Avenida Divina Pastora 7, 18012 Granada, Spain
              \and European Southern Observatory, Alonso de C\'{o}rdova 3107,Vitacura Casilla 7630355, Santiago, Chile}

   \date{Received .../ Accepted ....}

 \abstract
   {Messier 8 (M8), one of the brightest \hii\ regions in our Galaxy, is powered by massive O 
   type stars and is associated with recent and ongoing massive star formation. Two prominent massive star-forming regions associated with M8 are M8-Main, the particularly bright part of the large scale \hii\ region (mainly) ionised by the 
   stellar system Herschel 36 (Her~36) and M8~East (M8~E), which is mainly powered by a deeply embedded young stellar object (YSO), a bright infrared (IR) source, M8E-IR.}
   {We aim to study the interaction of the massive star-forming region M8~E with its surroundings using observations of assorted  diffuse and dense gas tracers that allow quantifying the kinetic temperatures and volume densities in this region. With a multi-wavelength view of M8~E, we want to investigate the cause of star formation. Moreover, we want to compare the star-forming environments of M8-Main and M8~E, based on their physical conditions and the abundances of the various observed species toward them.}
   {We used the Institut de Radioastronom\'{i}a Millim\'{e}trica (IRAM)~30~m telescope to perform an imaging spectroscopy survey of the $\sim1$~pc scale molecular environment of  M8E-IR and also performed  deep integrations toward the source itself. We imaged and analysed data for the  $J$ = 1 $\to$ 0 rotational transitions of $^{12}$CO, $^{13}$CO, N$_2$H$^+$, HCN, H$^{13}$CN, HCO$^+$, H$^{13}$CO$^+$, HNC and HN$^{13}$C observed for the first time toward M8~E. To visualise the distribution of the dense and diffuse gas in M8~E, we compare our velocity integrated intensity maps of $^{12}$CO, $^{13}$CO and N$_2$H$^+$ with ancillary data taken at infrared and sub-millimeter wavelengths. We used techniques both assuming Local Thermodynamic Equilibrium (LTE) and non-LTE to determine column densities of the observed species and to constrain the physical conditions of the gas responsible for their emission. Examining the Class 0/ I and Class II YSO populations in M8~E, allows us to explore the observed ionization front (IF) as seen in the high resolution Galactic Legacy Infrared Mid-Plane Survey Extraordinaire (GLIMPSE) 8~$\mu$m emission image. The difference between the ages of the YSOs and their distribution in M8~E were used to estimate the speed of the IF.} 
   {We find that $^{12}$CO probes the warm diffuse gas also traced by the GLIMPSE 8~$\mu$m emission, while N$_2$H$^+$ traces the cool and dense gas following the emission distribution of the APEX Telescope Large Area Survey of the Galaxy (ATLASGAL) 870~$\mu$m dust continuum. We find that the star-formation in M8~E appears to be triggered by the earlier formed stellar cluster NGC~6530, which powers an \hii\ region giving rise to an IF that is moving at a speed $\geq$ 0.26~km~s$^{-1}$ across M8~E. Based on our qualitative and quantitative analysis, we believe that the $J$ = 1 $\to$ 0 transition lines of N$_2$H$^+$ and HN$^{13}$C are more direct tracers of dense molecular gas than the $J$ = 1 $\to$ 0 transition lines of HCN and HCO$^+$. We derive temperatures of 80~K and 30~K for the warm and cool gas components, respectively, and constrain H$_2$ volume densities to be in the range of 10$^4$--10$^6$~cm$^{-3}$. Comparison of the observed abundances of various species reflects the fact that M8~E is at an earlier stage of massive star formation than M8-Main.} 
   {}

\titlerunning{Messier 8 east}
   \maketitle
%
\begin{figure*}[htp]
\centering
\subfigure{\includegraphics[width=183mm]{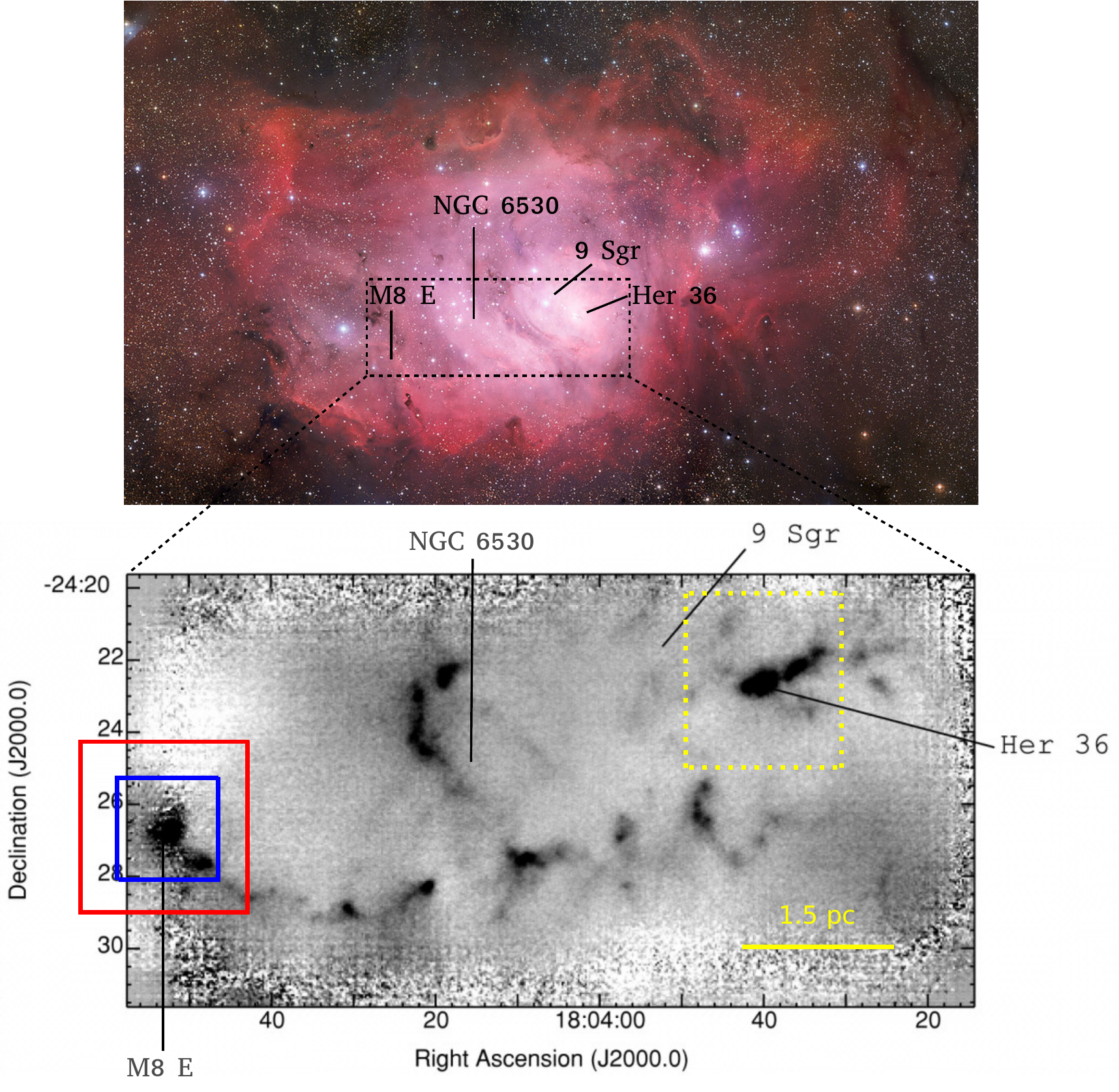}}

\caption{Upper panel: Image of M8 taken with Wide Field Imager (WFI) attached to the MPG/ESO 2.2~m telescope at the La Silla Observatory. The image is based on the maps observed by using three different broadband filters: B (at $\sim$ 456.252~nm), V (at 539.562~nm), R (at 651.725~nm) and one narrow-band filter: H~$\alpha$ (at 658.827~nm). Lower panel: Map of M8 at 450~$\mu$m adopted from \citet[Fig.~2]{2002ApJ...580..285T}.  The boundaries of the observed maps shown in Fig. \ref{mean_maps} are marked in red (Fig.3, top row) and blue (Fig. 3, middle and bottom rows). Our observed maps extend further to the east, beyond the edge of this image. The boundary of the observed maps of various species reported in \citet{refId0} is shown in yellow colored dots. In both panels, the positions of the luminous stellar systems Her~36 and 9~Sgr, which  both are located in the central optically bright region of M8, are marked, along with that of the open cluster NGC~6530 and the eastern star forming region M8-E.} 
  
\label{m8_overview} 
\end{figure*}

\section{Introduction\label{intro}}

Massive stars contribute significantly to the evolution of galaxies by injecting radiative and mechanical energy into the interstellar medium (ISM). This energy input stirs the environment around massive stars through stellar winds, ionization and heating of the gas, and through supernovae explosions, all processes that considerably change the chemical composition and structure of the ISM in their neighborhood \citep[for overviews, see][]{ 2010pcim.book.....T, Tielens2013, 2011piim.book.....D}. Hence, the interaction of massive stars with their surroundings can affect the star formation process either by quenching star formation or by triggering it: formation of cloud and intercloud phases in the ISM, leading to disruption of molecular clouds, can result in halting star formation, while compression of the surrounding gas or the propagation of photoionisation induced shocks can set off star formation, a scenario originating with \citet{Elmegreen1977}; see, e.g.,  \citet{2007A&A...467.1125U} and \citet{2013ApJ...776....1K} for recent observational studies. Massive stars give rise to bright \hii\ regions and photodissociation regions (PDRs). \hii\ regions comprise hot ionized gas irradiated by strong ultraviolet (UV, h$\nu$ $>$ 13.6~eV) radiation from one or more nearby hot luminous stars. PDRs are at the interface of these \hii\ regions and the cool molecular cloud shielded from UV radiation from the illuminating star \citep{1999RvMP...71..173H}. In PDRs,  the thermal and chemical processes are regulated by far-UV (FUV, 6~eV $<$ h$\nu$ $<$ 13.6~eV) photons. In order to understand how the ISM gets affected by the interaction with massive stars, it is of fundamental importance to study massive star-forming regions.\

\begin{table*}
\label{lines}
\centering
\begin{threeparttable}
\caption{Molecular lines mapped in the M8~E region.}
\begin{tabular}{c c c c c c c c c}
\hline\hline
\noalign{\smallskip}
 Species/Line & $\nu$ & $E_{\rm up}$ & $\theta$ & $\int T_{\rm MB} d\varv$ (peak) & $\varv_{\rm LSR}$ & rms (K~km~s$^{-1}$) & $n_{\rm crit}$\tnote{a}~ & Size\\
 \hline
 \noalign{\smallskip}
& (GHz) & (K) & ($\arcsec$) & (K~km~s$^{-1}$) & (km~s$^{-1}$) & (K~km~s$^{-1}$) & (cm$^{-3}$) & ($\arcsec \times \arcsec$)\\
 \hline
 \noalign{\smallskip}
$^{12}$CO (1 $\to$ 0) & 115.2712 & 5.5 & 22.5 & 440.2 & 10.5 & 0.71 & 2.2 $\times$ 10$^3$ & 260 $\times$ 260\\
$^{13}$CO (1 $\to$ 0) & 110.2013 & 5.3 & 23.5 & 86.9 & 9.9 & 0.33 & 1.9 $\times$ 10$^3$ & 260 $\times$ 260\\
N$_2$H$^{+}$ (1 $\to$ 0) & 93.1733 & 4.5 & 27.8 & 20.3 & 10.7 & 0.21 & 3.7 $\times$ 10$^5$ & 260 $\times$ 260\\
HCN (1 $\to$ 0) & 88.6316 & 4.3 & 29.3 & 77.7 & 10.6 & 0.17 & 1 $\times$ 10$^6$ & 160 $\times$ 160\\
H$^{13}$CN (1 $\to$ 0) & 86.3399 & 4.1 & 30.0 & 9.0 & 10.8 & 0.12 & & 160 $\times$ 160\\
HCO$^{+}$ (1 $\to$ 0) & 89.1885 & 4.3 & 29.1 & 42.6 & 10.8 & 0.21 &1.9 $\times$ 10$^5$ & 160 $\times$ 160\\
H$^{13}$CO$^{+}$ (1 $\to$ 0) & 86.7542 & 4.2 & 30.0 & 3.8 & 10.7 & 0.16 & 1.8 $\times$ 10$^5$ & 160 $\times$ 160 \\
HNC (1 $\to$ 0) & 90.6635 & 4.4 & 28.6 & 25.4 & 10.8 & 0.21 & 3.2 $\times$ 10$^5$ & 160 $\times$ 160  \\
HN$^{13}$C (1 $\to$ 0) & 87.0908 & 4.2 & 30.0 & 1.9 & 10.8 & 0.17 & & 160 $\times$ 160\\

 \hline
\end{tabular}

\small Notes: Columns are, left to right, species/line, frequency, upper level energy, FWHM beam size, peak integrated flux density in the map, centroid LSR velocity, rms noise, critical density for optically thin emission and map size. 

The measured line parameters (flux, centroid LSR velocity and rms) are for the central position, which corresponds to that of M8E-IR and is given in Section~2.

\begin{tablenotes}

\item[a]\small Critical densities calculated for species whose collisional rate coefficients are provided by LAMDA. For $^{12}$CO and $^{13}$CO, the coefficients are determined at 80~K and for N$_2$H$^{+}$, HCN, HCO$^{+}$, H$^{13}$CO$^{+}$ and HNC, the coefficients are determined at 30~K.   
\end{tablenotes}

 \label{all_obs}
 \end{threeparttable}
\end{table*}

Messier 8 (M8), the Lagoon Nebula, is one of the most prominent \hii\ regions in our galaxy. Associated with is central region is a well-developed PDR \citep{refId0} that represents the second most intense $^{12}$CO emission source known (in 1996) \citep{1997A&A...323..529W}. Located  in the Sagittarius-Carina arm, near our line of sight toward the Galactic Center, it is relatively close to the Sun, at a distance of $\sim$ 1.25~kpc (1$\arcmin$ corresponding to 0.36~pc) (\citealt{2004ApJ...608..781D} and \citealt{2006MNRAS.366..739A}). Recently, this distance has been confirmed by Gaia parallaxes that yield a value of 1.35 kpc with an uncertainty of 9\% \citep{Damiani2019}.
M8 (Fig.~\ref{m8_overview}) as a whole is an extended  \hii\ region ($\sim$ 10~pc across) that is powered by the open stellar cluster NGC~6530 \citep{Prisinzano2005}, which contains several O-type stars,  and by the spectroscopic binary 9 Sagittarii (9 Sgr) consisting of an O3.5 V and an O5-5.5 V star \citep{Rauw2012}. M8 is also associated with two major regions with more recent star formation activity: one near its brightest part, the Hourglass Nebula (hereafter M8-Main) that is illuminated by the bright and very young multiple stellar system Herschel 36 (Her 36) (O7.5 V +
O9 V + B0.5 V \citep{Arias2010}) and a region to the south-east, M8 east (M8~E), where at least two massive stars have very recently formed, and whose dust continuum emission rivals in brightness that of M8-Main at far-infrared (FIR) and submillimeter wavelengths \citep{2008hsf2.book..533T}.

Using the Stratospheric Observatory for Infrared Astronomy (SOFIA, \citealt{2012ApJ...749L..17Y}), the Atacama Pathfinder EXperiment 12 meter submilimeter telescope (APEX, \citealt{2006SPIE.6267E..14G}) and the 30~meter millimeter telescope operated by the Institut de Radioastronom\'{i}a Millim\'{e}trica (IRAM\footnote{IRAM is supported by INSU/CNRS, the MPG (Germany), and IGN (Spain)}) on Pico Veleta in Spain, we performed an extensive survey of M8-Main, as reported in two recent publications:  We described the morphology of the volume around Her 36 and determined the physical conditions of the gas surrounding it \citep{refId0} and shed some light on the formation process of hydrocarbons in M8-Main, which is a high UV flux PDR with $G_{\rm 0}$ $\sim$ 10$^5$ in Habing units \citep{2019A&A...626A..28T}. 

\begin{figure}[h!]

\subfigure{\includegraphics[width=80mm]{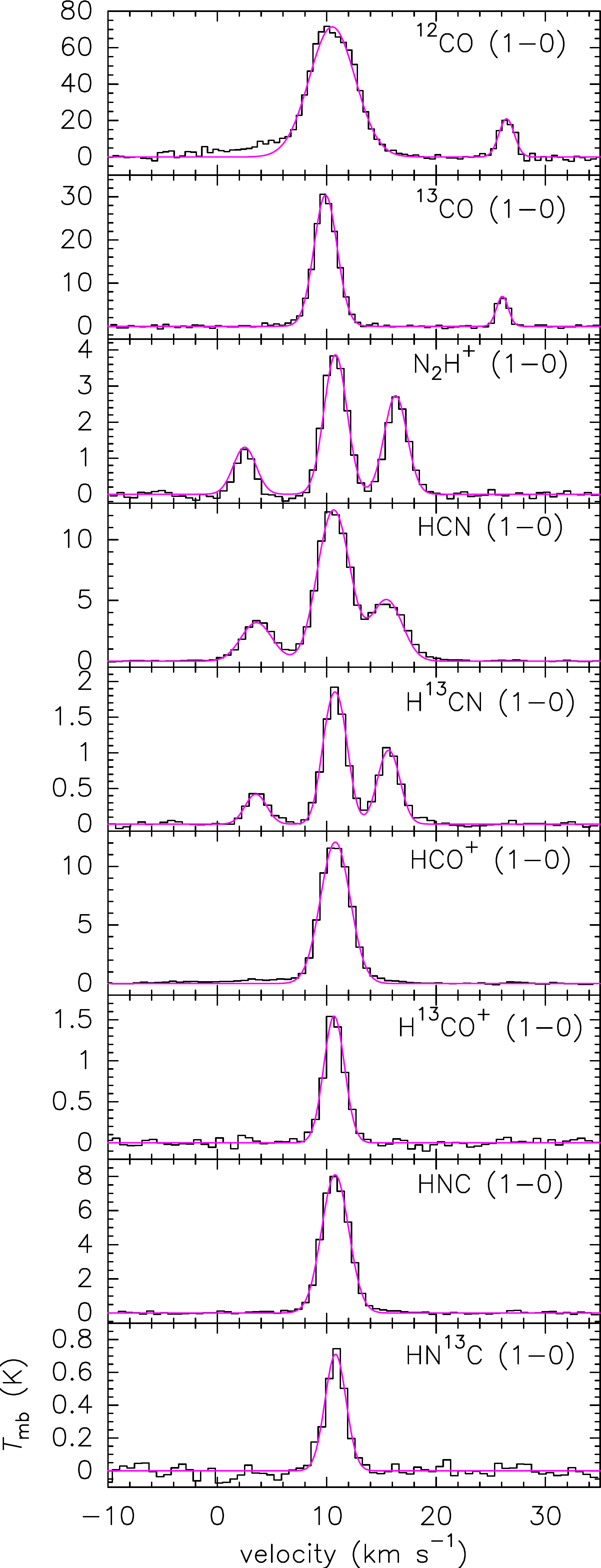}}

\caption{Spectra of the species listed in Table~\ref{all_obs}, observed toward M8E-IR (black) with Gaussian fits (in pink), to a single line and, for the NH$_2$H$^+$, HCN and H$^{13}$13CN lines to their hfs components.} 
  
\label{all_spec} 
\end{figure}

M8~E is a massive star-forming region about 5~pc in projected distance away from M8-Main and it lies within a  molecular cloud with a size of a few arcmin (roughly 1~pc) \citep{2008hsf2.book..533T}. A small ($<$ 0.13~pc$^2$ $\sim$ 1 arcmin$^2$) region contains a quite rich embedded cluster comprising 7 IR sources that is associated with the molecular gas. It includes a Zero Age Main Sequence (ZAMS) B2 star, powering a very small ($0\rlap{}.''6$ $\sim$ 750~AU in diameter) ultracompact \hii\ region, known as M8E-Radio, and, only $\sim$ 7$''$ (0.04~pc) away, a massive young stellar object (YSO), M8E-IR, which is likely to become a B0 star and dominates the cluster's near and mid infrared IR radiation up to a wavelength of $\sim$ 24~$\mu$m \citep{2008JPhCS.131a2024L}. \
In the past, (sub)mm wavelength imaging of M8~E 
has targeted dust continuum emission at 1300, 850 and 450 $\mu$m and resulted in the detection  a number of cores and also CO emission 
\citep{2002ApJ...580..285T, 2005ApJ...625..864Z}.
In particular, high-velocity molecular gas tracing  a molecular outflow  was discovered to originate in M8E-IR by mapping the $^{12}$CO $J$ = 2$\to$ 1 transition \citep{Mitchell1992,2005ApJ...625..864Z}.  

 


Quite surprisingly, the literature on observations of M8~E in lines from molecules other than CO is quite sparse. The goal of this study was to collect and analyze data from a variety of molecular lines to establish an inventory of the observed diffuse and dense gas tracers  
that can be used to  investigate the physical and chemical conditions in the M8~E massive star-forming region and compare this molecular environment with that of 
M8-Main. Our observations were done in the 3~mm band using the Eight MIxer Receiver (EMIR\footnote{http://www.iram.es/IRAMES/mainWiki/EmirforAstronomers}, \citealt{2012A&A...538A..89C}) on the IRAM~30~m telescope. 
In this paper, we present observations of various diffuse and dense gas tracers found toward M8~E that consisted of a mapping survey with different frequency coverage of a large (260$\arcsec$ $\times$ 260$\arcsec$) and a smaller (160$\arcsec$ $\times$ 160$\arcsec$) region ($\approx 1.6~\rm{pc}\times1.6~\rm{pc}$ and $1~\rm{pc}\times 1~\rm{pc}$, respectively) centered on M8E-IR. The observations are described in Section~2, and velocity integrated intensity maps of the various observed species are presented in Section~3. The data is analysed quantitatively in Section~4. Finally, we discuss the results and summarise the main conclusions of this work in Sections~5 and 6 and give a brief outlook on further studies of M8's interesting environment in Section 7.

\section{Observations}
We used the  IRAM 30~m telescope in 2016 August for observations of M8~E covering most of the 3~mm atmospheric window. We employed EMIR as a frontend, followed by a Fast Fourier Transform Spectrometer (FFTS) as a backend. The FFTS channel spacing of 200 kHz corresponds to a  velocity resolutions of 0.70 and 0.52 km~s$^{-1}$ at the frequencies of the lowest and highest frequency lines under discussion here, 85.4 and 115.3 GHz, respectively. 
In this paper we focus on the $J$ = 1 $\to$ 0 transitions of $^{12}$CO, $^{13}$CO, N$_2$H$^+$, HCN, H$^{13}$CN, HCO$^+$, H$^{13}$CO$^+$, HNC and HN$^{13}$C and on the $J$ = 5 $\to$ 4 and $J$ = 6 $\to 5$ multiple $K$ series of CH$_3$CCH. Information on the observed $J$ = 1 $\to$ 0 lines is given in 
Table \ref{lines}, while Table \ref{ch3cch_table} summarizes the CH$_3$CCH lines. The full molecular inventory of M8~E will be discussed in a forthcoming publication. 

Mapping observations in On-The-Fly (OTF) total power mode were centered on M8E-IR at, respectively, right ascension and declination,  ($\alpha$,$\delta$)$_{\rm J2000}$ = 18$^{\rm h}$04$^{\rm m}$53.$^{\rm s}$3, -24$^\circ$26$\arcmin$42.$''$3.
\citep{2008hsf2.book..533T}. The $^{12}$CO, $^{13}$CO and N$_2$H$^+$ line emission maps have a size of $\sim$ 260$''$ $\times$ 260$''$, while the HCN, H$^{13}$CN, HCO$^+$, H$^{13}$CO$^+$, HNC and HN$^{13}$C line maps have a size of $\sim$ 160$''$ $\times$ 160$''$. Each subscan lasted 25~s and the integration time on the off-source reference position was 5~s. The offset position relative to the center was at (30$\arcmin$, -30$\arcmin$) and the pointing accuracy ($<$ 3$\arcsec$) was maintained by pointing at the bright calibrator, the QSO B1757$-$240, every 1$-$1.5~hrs. We also performed pointed observations with deep integrations toward the M8E-IR position providing us with the better S/N (signal to noise ratio) required for our spectral analysis and from which the CH$_3$CCH data profit. A forward coupling efficiency, $\eta_{\rm f}$, of 0.94 and a main beam efficiency, $\eta_{\rm MB}$, of 0.7 
were adopted for the 3 mm module of the EMIR receiver. These efficiencies are defined and their values reported on the IRAM30m efficiencies website\footnote{http://www.iram.es/IRAMES/mainWiki/Iram30mEfficiencies} and in the (2015) IRAM commissioning report also linked in the above mentioned website.

The reduction of the calibrated data to produce spectra and maps shown throughout the paper was performed using the Continuum and Line Analysis Single dish Software (CLASS) and the GREnoble Graphic (GREG) softwares that are a part of the Grenoble Image and Line Data Analysis Software (GILDAS\footnote{www.iram.fr/IRAMFR/GILDAS/}, \citealt{2005sf2a.conf..721P}) package. A part of the data analysis was done using matplotlib \citep{Hunter:2007}, which is a python library. All observations are summarized in Tables~\ref{all_obs} and~\ref{ch3cch_table}.\

\section{Results}

\subsection{Observed spectra toward M8E-IR}
Figure~\ref{all_spec} shows the spectra of the species listed in Table~\ref{all_obs}, observed along the brightest line-of-sight (LOS), i.e. toward M8E-IR. The $^{12}$CO and $^{13}$CO line parameters were derived from two-component Gaussian fits, where initial guesses of 5--20~km~s$^{-1}$ and 24--30~km~s$^{-1}$ were provided. For HCO$^+$, H$^{13}$CO$^+$, HNC and HN$^{13}$C line parameters were derived from single-component Gaussian fit. The line parameters of N$_2$H$^+$, HCN and H$^{13}$CN were obtained from fits that considered these molecules' hyperfine structure (hfs). 
The derived centroid velocities are listed in Table~\ref{all_obs}. For $^{12}$CO and $^{13}$CO, the centroid velocity of the brightest component is mentioned. Similarly, for N$_2$H$^+$, HCN and H$^{13}$CN, the centroid velocity of the brightest hfs component is given.


The systemic LSR velocity of M8~E thus determined is $\approx$ +11~km~s$^{-1}$, which is similar to the $ +6$ to $+11$~km~s$^{-1}$, found for  M8-Main \citep{refId0}. This similarity clearly indicates that both regions are in the same complex, although at different stages of their evolution. 
In addition, we see in both the $^{12}$CO and $^{13}$CO lines emission at a very different LSR velocity, $\sim$ +26~km~s$^{-1}$, which corresponds to a different molecular cloud along the LOS to M8 that is unrelated to this region \citep{2001ApJ...547..792D}.  This other velocity component was also reported by \citet{1992ApJ...386..604M} in their $^{12}$CO $J$ = 2 $\to$ 1 spectrum. 
Velocity integrated intensity maps of the $J$ = 1 $\to$ 0 transition of $^{12}$CO and $^{13}$CO from the +26~km~s$^{-1}$ cloud are shown in Fig.~\ref{v26}.

\begin{figure*}[htp]
\centering
\subfigure{\includegraphics[width=55mm]{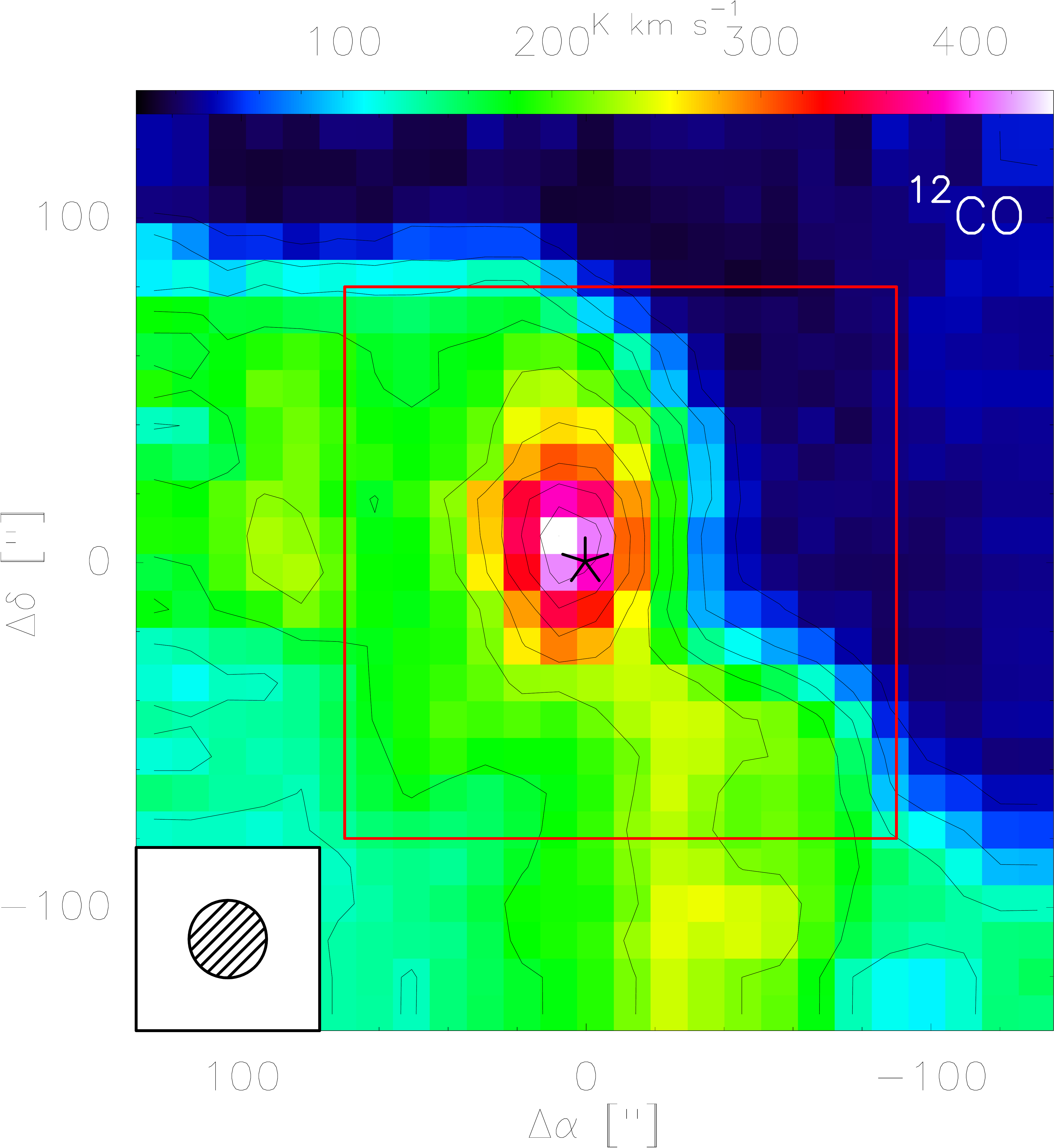}}\quad
\subfigure{\includegraphics[width=55mm]{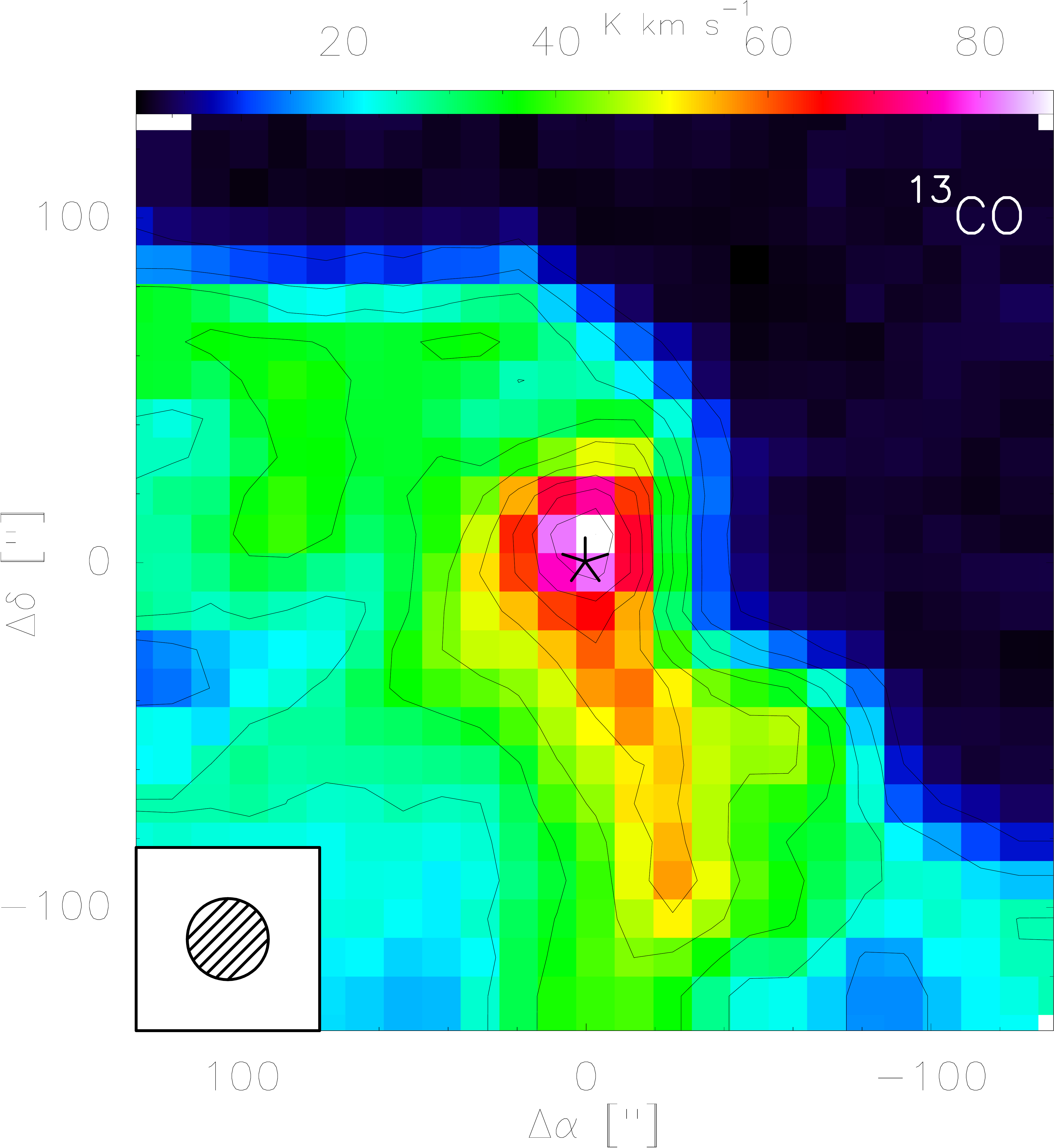}}\quad
\subfigure{\includegraphics[width=55mm]{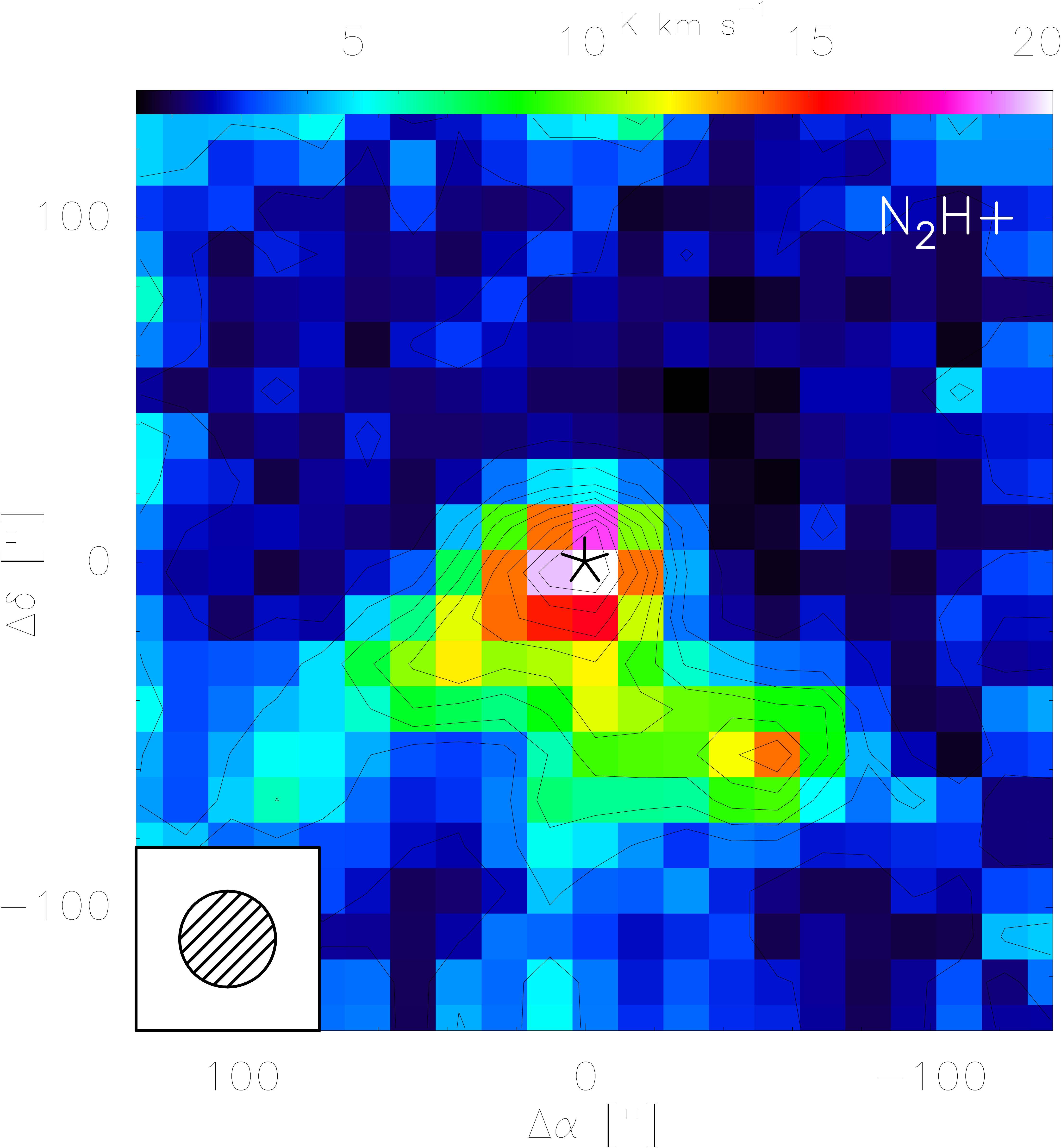}}
\subfigure{\includegraphics[width=55mm]{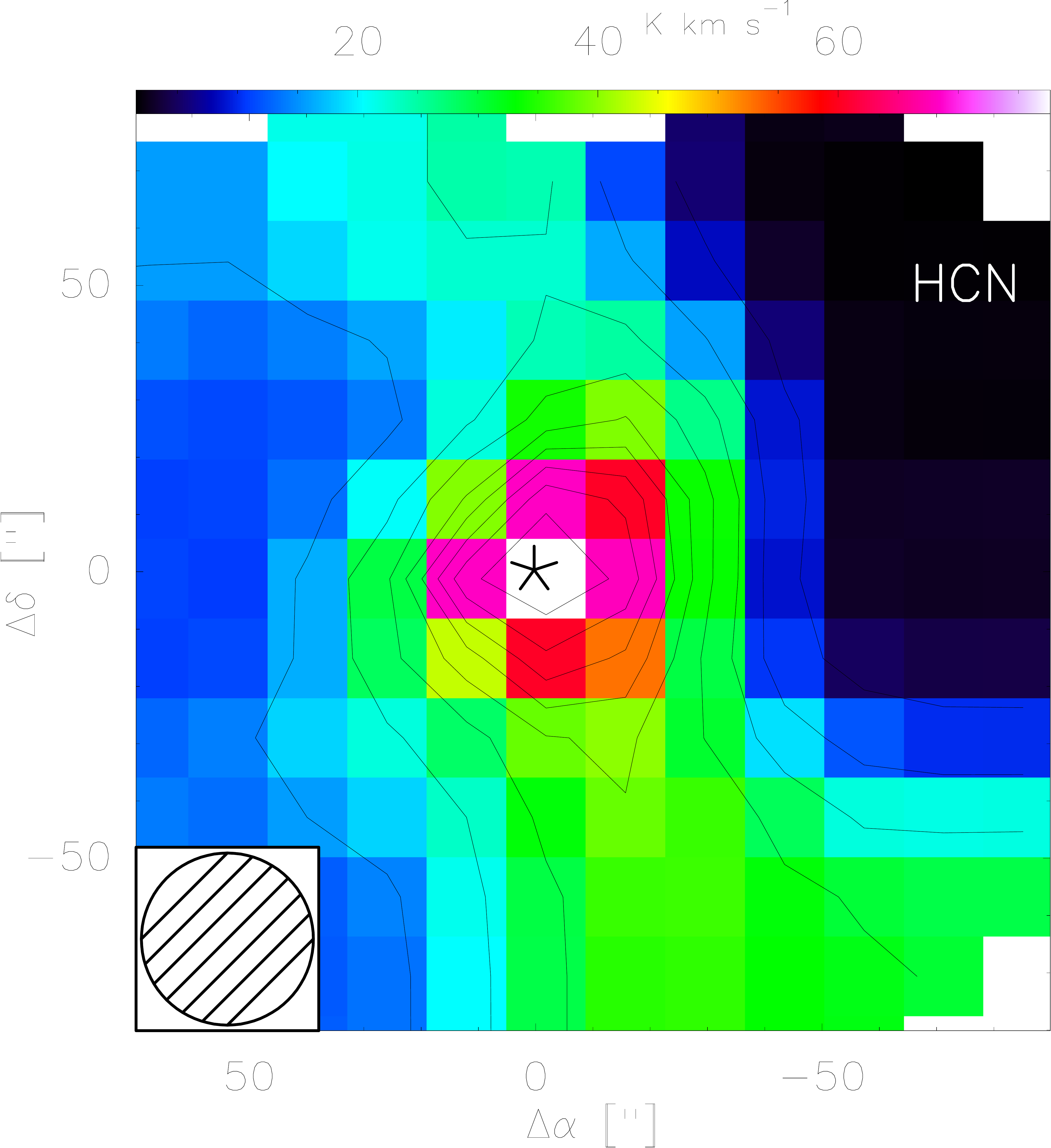}}\quad
\subfigure{\includegraphics[width=55mm]{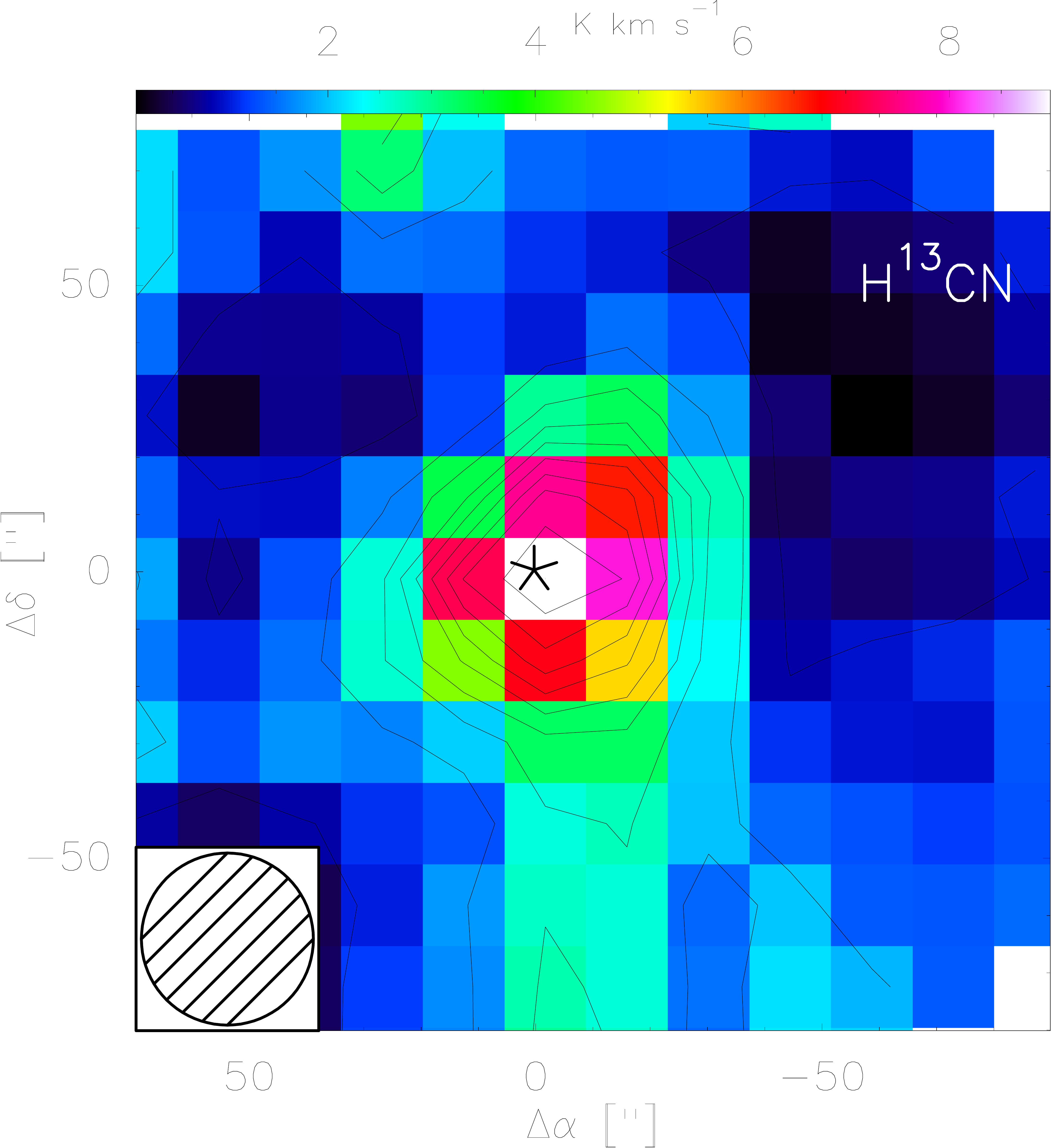}}\quad
\subfigure{\includegraphics[width=55mm]{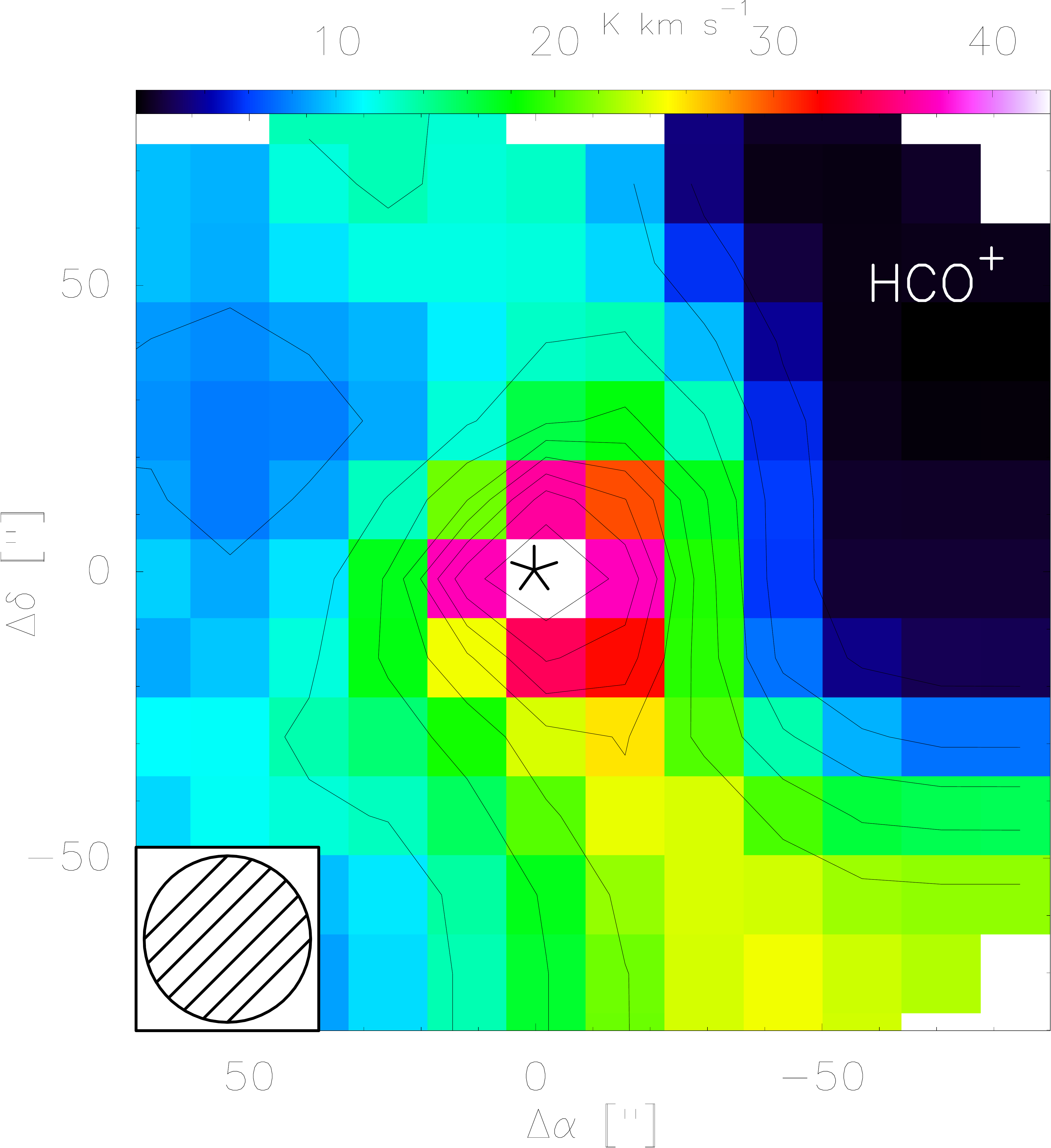}}
\subfigure{\includegraphics[width=55mm]{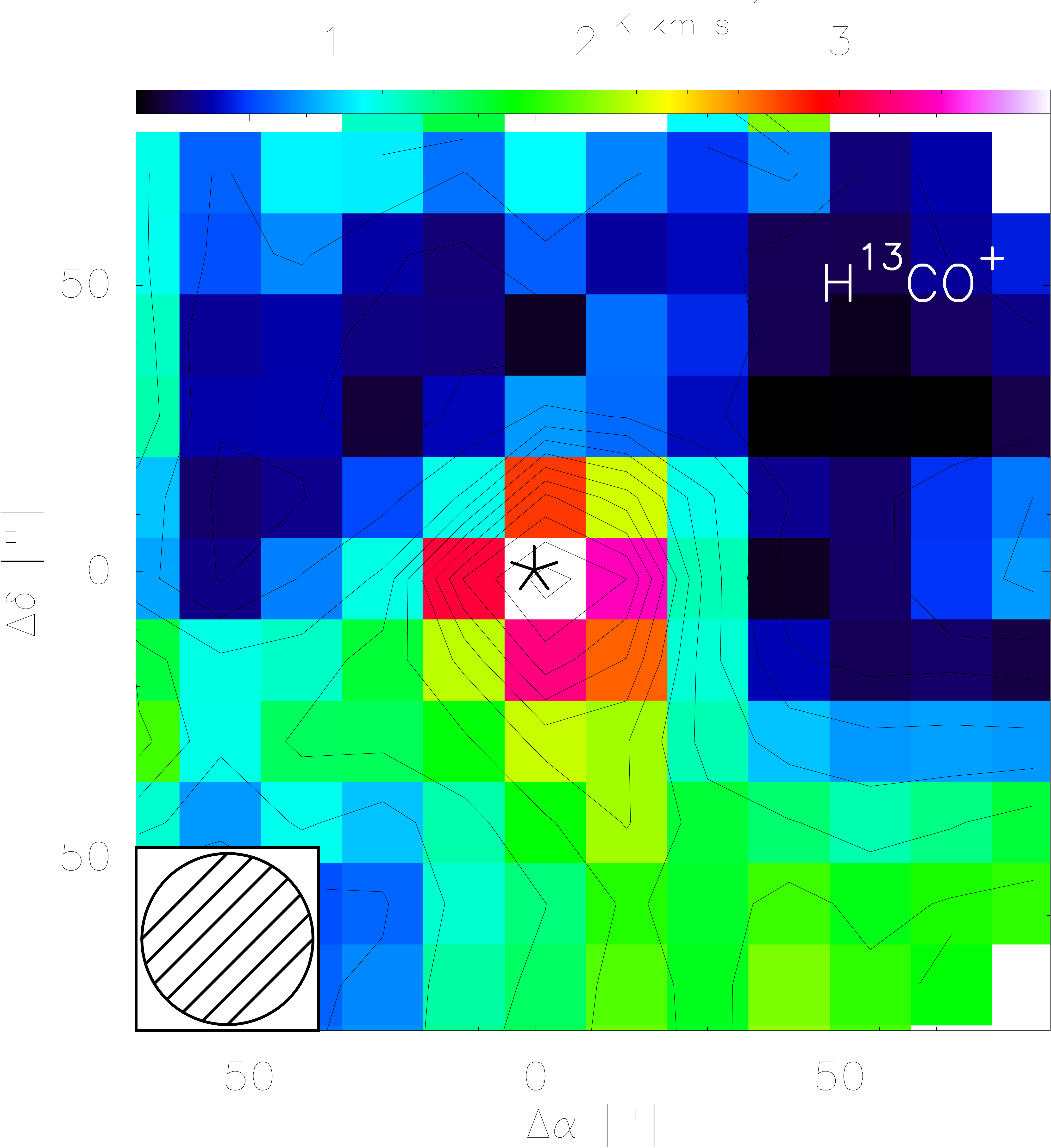}}\quad
\subfigure{\includegraphics[width=55mm]{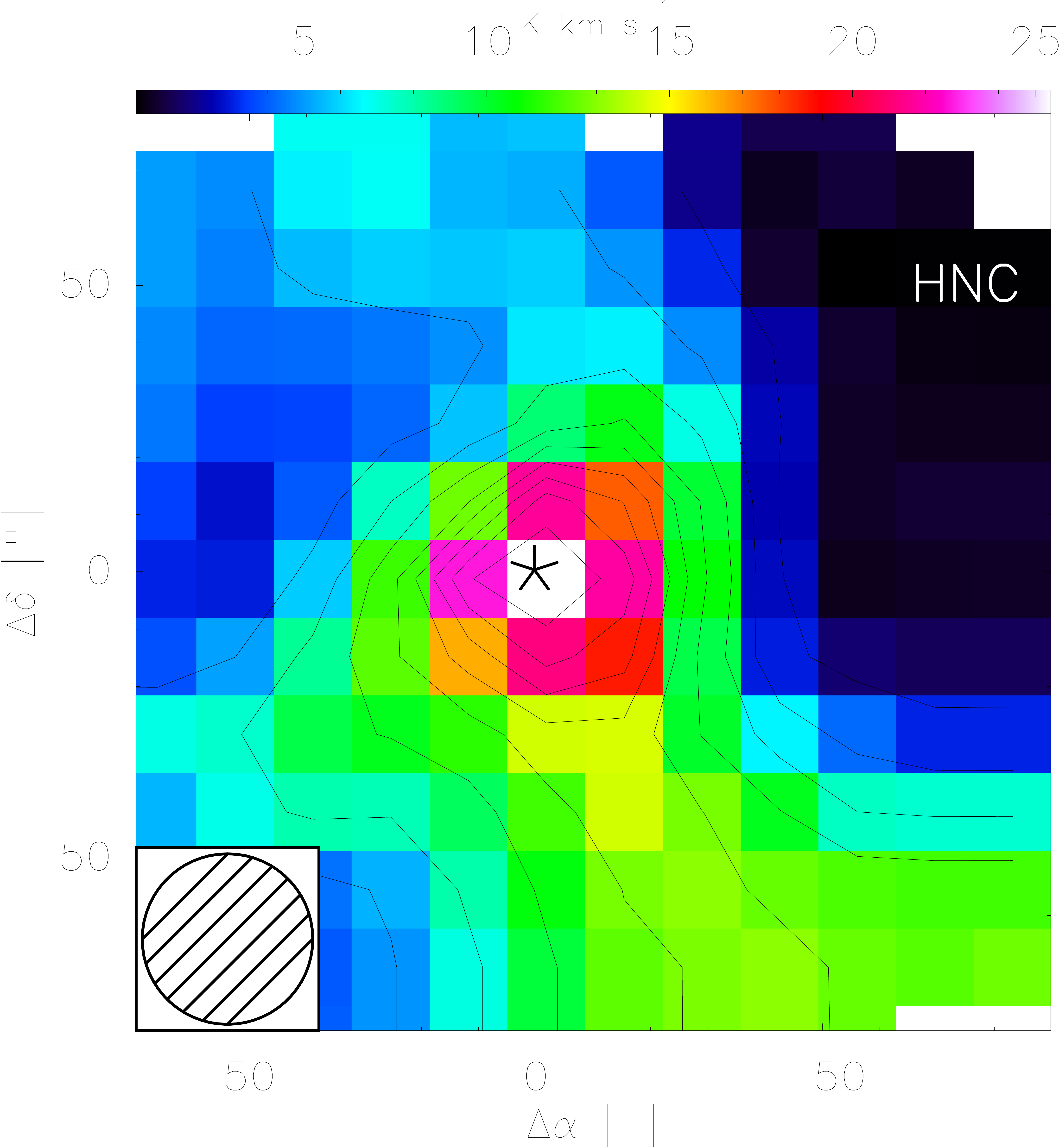}}\quad
\subfigure{\includegraphics[width=55mm]{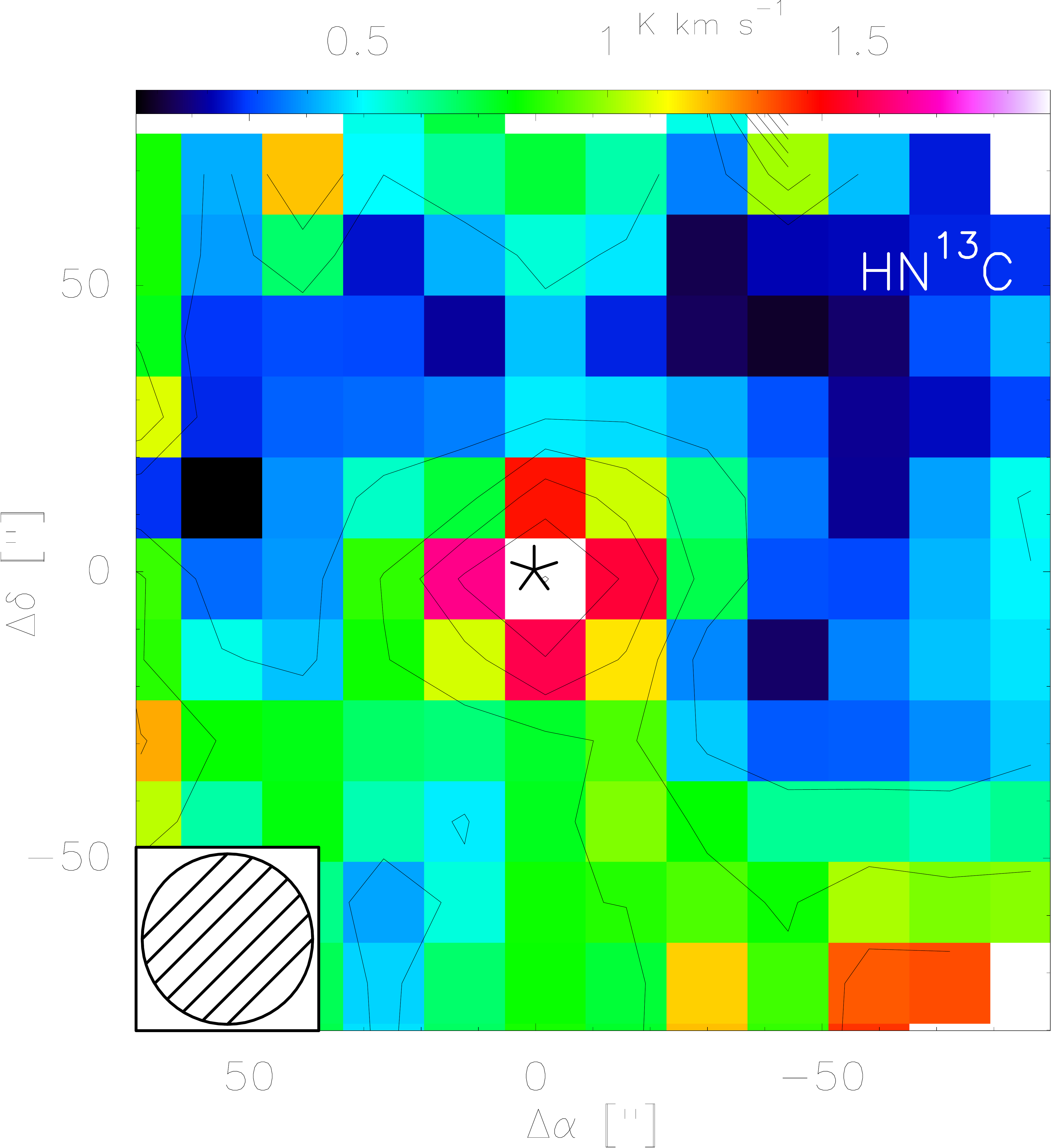}}
  \caption{Color maps of the velocity integrated intensity of the $J$ = 1 $\to$ 0 transition of $^{12}$CO, $^{13}$CO, N$_2$H$^+$, HCN, H$^{13}$CN, HCO$^{+}$, H$^{13}$CO$^{+}$, HNC and HN$^{13}$C toward M8~E. The square outlined by the red dashed lines in the map of $^{12}$CO denotes the area of the smaller maps of HCN, H$^{13}$CN, HCO$^{+}$, H$^{13}$CO$^{+}$, HNC and HN$^{13}$C. Position offsets are relative to the position of M8E-IR (marked with an asterisk) and given in Section~2. The contour levels are from 10\% to 100\% of the corresponding peak emission (given in Table~\ref{all_obs}) for $^{12}$CO, $^{13}$CO, N$_2$H$^+$, HCN, H$^{13}$CN, HCO$^{+}$ and HNC. For H$^{13}$CO$^{+}$ and HN$^{13}$C maps, the contour levels are from 3rms to peak emission in steps of 2rms. All maps are plotted using original FWHM beam sizes shown in the lower left of each map and listed in Table~\ref{all_obs}.}

 \label{mean_maps} 
\end{figure*}

\begin{figure}[htp]
\subfigure{\includegraphics[width=82mm]{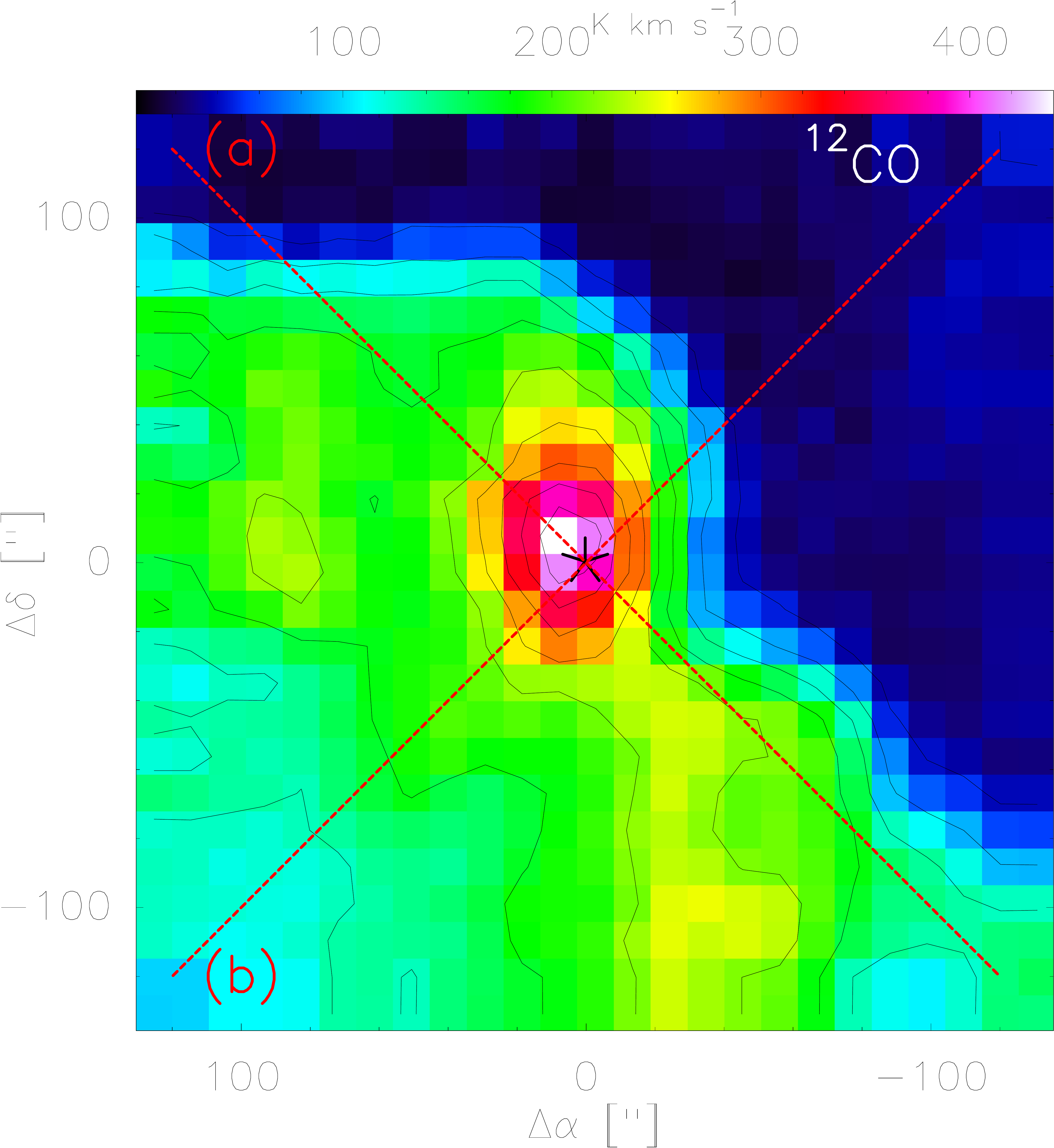}}

\caption{Velocity integrated intensity map of $^{12}$CO marked with two red dashed cuts: (a) going from (120$''$, 120$''$) to (-120$''$, -120$''$) and (b) going from (120$''$, -120$''$) to (-120$''$, 120$''$). Intensity profiles of various species are plotted along these directions in Fig.~\ref{int_strips}.} 
  
\label{strips_cuts} 
\end{figure}

\begin{figure*}[h!]

\subfigure{\includegraphics[width=85mm]{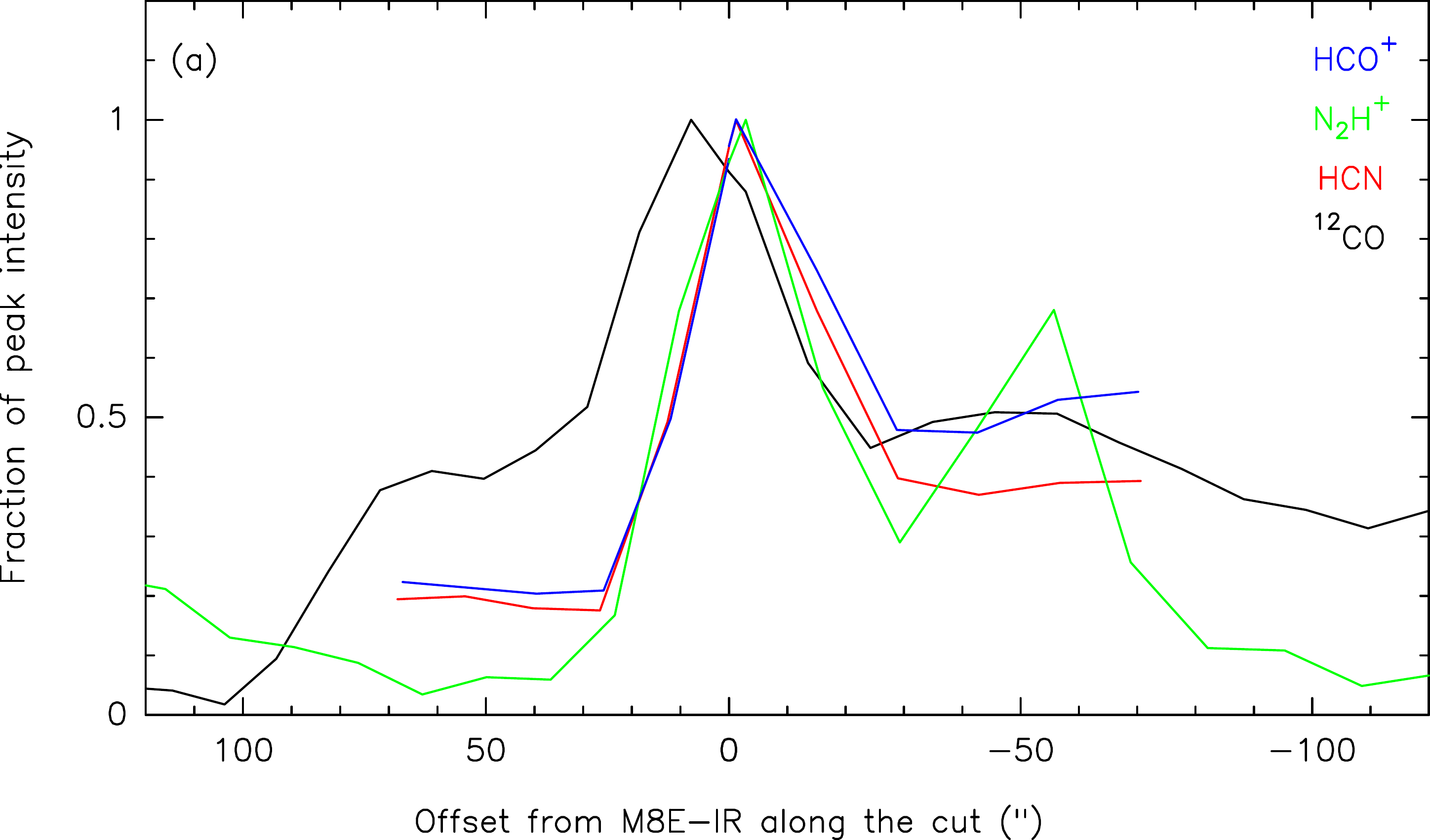}}\quad\quad 
\subfigure{\includegraphics[width=85mm]{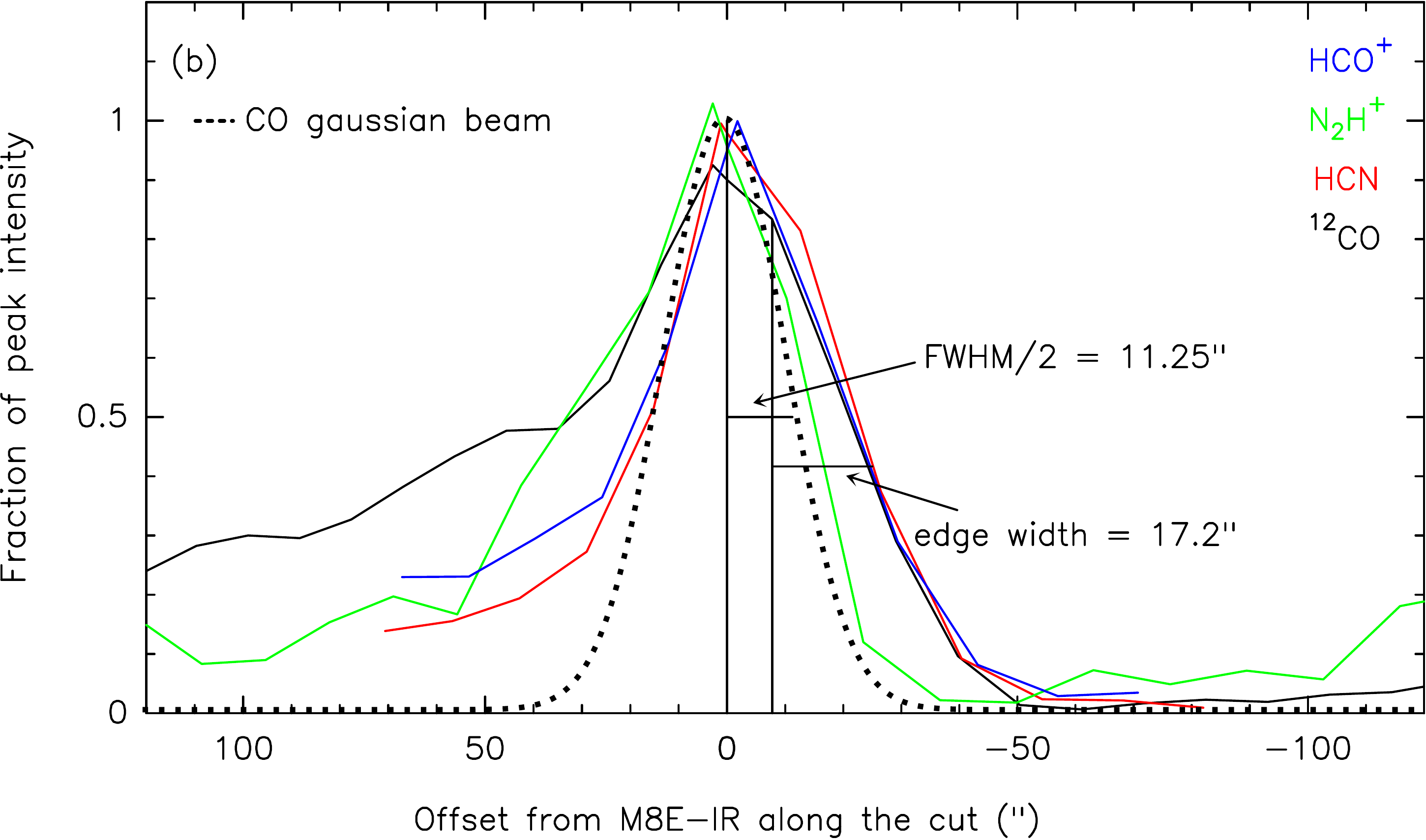}}

\caption{Velocity integrated intensities normalized to their value at peak vs offset ($''$) from M8E-IR along the cuts shown in Fig.~\ref{strips_cuts}.} 
  
\label{int_strips} 
\end{figure*}

\subsection{Spatial distribution of the molecular line emission}

Table~\ref{all_obs} lists, for all the observed molecular lines, the line parameters determined from our spectra, which include the emission maxima and rms noise as obtained from the velocity integrated intensity maps and the calculated critical densities (for optically thin emission).  Figure~\ref{mean_maps} shows velocity integrated intensity maps of the $J$ = 1 $\to$ 0 transitions of $^{12}$CO, $^{13}$CO, N$_2$H$^+$, HCN, H$^{13}$CN, HCO$^+$, H$^{13}$CO$^+$, HNC and HN$^{13}$C. The intensities of the N$_2$H$^+$ and HCN lines were integrated over a velocity range of 0 to 20~km~s$^{-1}$ and -5 to 20~km~s$^{-1}$, respectively in order to cover these lines' hfs components. The emission intensities from all molecules peak at or very close to the position of M8E-IR, 
As one would expect, the distribution of the $^{12}$CO emission is spread out the most, while the regions showing  H$^{13}$CN and N$_2$H$^+$ emission are the most concentrated. To quantify the extent of the emission regions, we have determined their angular sizes by 
fitting two-dimensional (2D) gaussians to the velocity integrated intensity maps, such that a contour plot at a cut through the Gaussian will be an ellipse. We used the "run gauss\_2d" command in GILDAS software for this. The elliptical emission size can be derived by defining boundaries of emission in our intensity maps and by guessing input values for the peak intensity and its corresponding position. The fit resulted in an angular size of the emitting region of H$^{13}$CN of $\sim$ 40$\arcsec$ and for that of  N$_2$H$^+$ of $\sim$ 72$\arcsec$. The $^{12}$CO and $^{13}$CO emission is extending over a larger area to the south-west and north-east of M8E-IR. Molecules with high critical densities [$\sim$ 10$^4$--10$^5$~cm$^{-3}$ \citep{2015PASP..127..299S}] such as HCN, HCO$^+$ and HNC, also show this extended emission but more pronounced in the south-west direction than to the  north-east. In contrast, only the  south-west branch is seen in the lines of these molecules' isotopologues, which have similar critical densities. In addition to the brightest emission toward M8E-IR, the intensity distribution of N$_2$H$^+$ shows a secondary peak south-west of M8E-IR ($\Delta\alpha$ = --55$\arcsec$, $\Delta\delta$ = --55$\arcsec$). 

Due to their high spontaneous emission rates, resulting in large critical densities, generally, the $J$ = 1 $\to$ 0 transition of the HCN, HCO$^+$ and HNC lines are considered to trace dense molecular gas. However, recent studies have shown that the $J$ = 1 $\to$ 0 transition of N$_2$H$^+$ is a more reliable dense gas tracer compared to the same transition of HCN, HCO$^+$ and HNC (\citealt{2017A&A...605L...5K}, \citealt{2017A&A...599A..98P} and \citealt{2020arXiv200306842B}). Despite having high critical densities, the $J$ = 1 $\to$ 0 transitions of HCN, HCO$^+$ and HNC (and other species) were observed in the Giant Molecular Clouds (GMCs) Orion A and B and W49 from relatively low density gas ($n_{\rm H}$ $\sim$ 10$^3$~cm$^{-3}$); thus the gas responsible for the emission of these species is not necessarily cool or dense  (\citealt{2017A&A...605L...5K}, \citealt{2017A&A...599A..98P} \& \citealt{Barnes2020}). Also, \citet{2017ApJ...841...25G} reported that if the electron abundance is of the order $X(\rm e^{-})$ $>$ 10$^{-5}$ or higher, HCN can be excited by collisions with electrons in regions with H$_2$ densities much lower than the mentioned critical density. 

Moreover, we see that the emission distribution of HN$^{13}$C follows that of N$_2$H$^+$ in M8~E, similar to what is found for OMC-1 by \citet{2019PASJ..tmp...32N}. Both N$_2$H$^+$ and HN$^{13}$C probe gas with very high visual extinction $A_{\rm v}$ $>$ 35 mag \citep{2017A&A...599A..98P}, hence tracing dust-embedded cores in molecular clouds. 

The emission intensity of all molecules decreases dramatically to the north (from $\sim$ $\Delta\delta$ = 100$\arcsec$, visible in $^{12}$CO and $^{13}$CO maps) and north-west (from $\sim$ $\Delta\alpha$ = 40$\arcsec$, visible in all maps) of M8E-IR. This is because M8~E is located at the edge of the compressed and warm molecular cloud that is strongly confined by the central \hii\ region of M8 \citep{2002ApJ...580..285T}. In order to visualize the presence of this steep edge, we plotted the emission distribution profiles of $^{12}$CO, HCN, HCO$^+$ and N$_2$H$^+$ along two cuts, which are shown in Fig.~\ref{strips_cuts}. We compared the progression of the intensities of various species along these cuts in Fig.~\ref{int_strips}. It can be seen that while moving along  cut (a), the emission of all species peaks at an offset 0$''$ (corresponding to the position of M8E-IR) and 
decreases gradually toward the east and west of it. But as we move along  cut (b), we see a sudden drop in the intensities 
away from M8E-IR to the  north-west. The observed width of this edge, $\sim$ 17$''$, is a little larger than the half width  ($0.5 \times$FWHM) of the CO beam as shown in Fig.~\ref{int_strips}~(b). So, we are able to resolve this edge, but only slightly. \\







\subsection{Ancillary data} 
\begin{figure*}[htp]
\subfigure{\includegraphics[width=80mm]{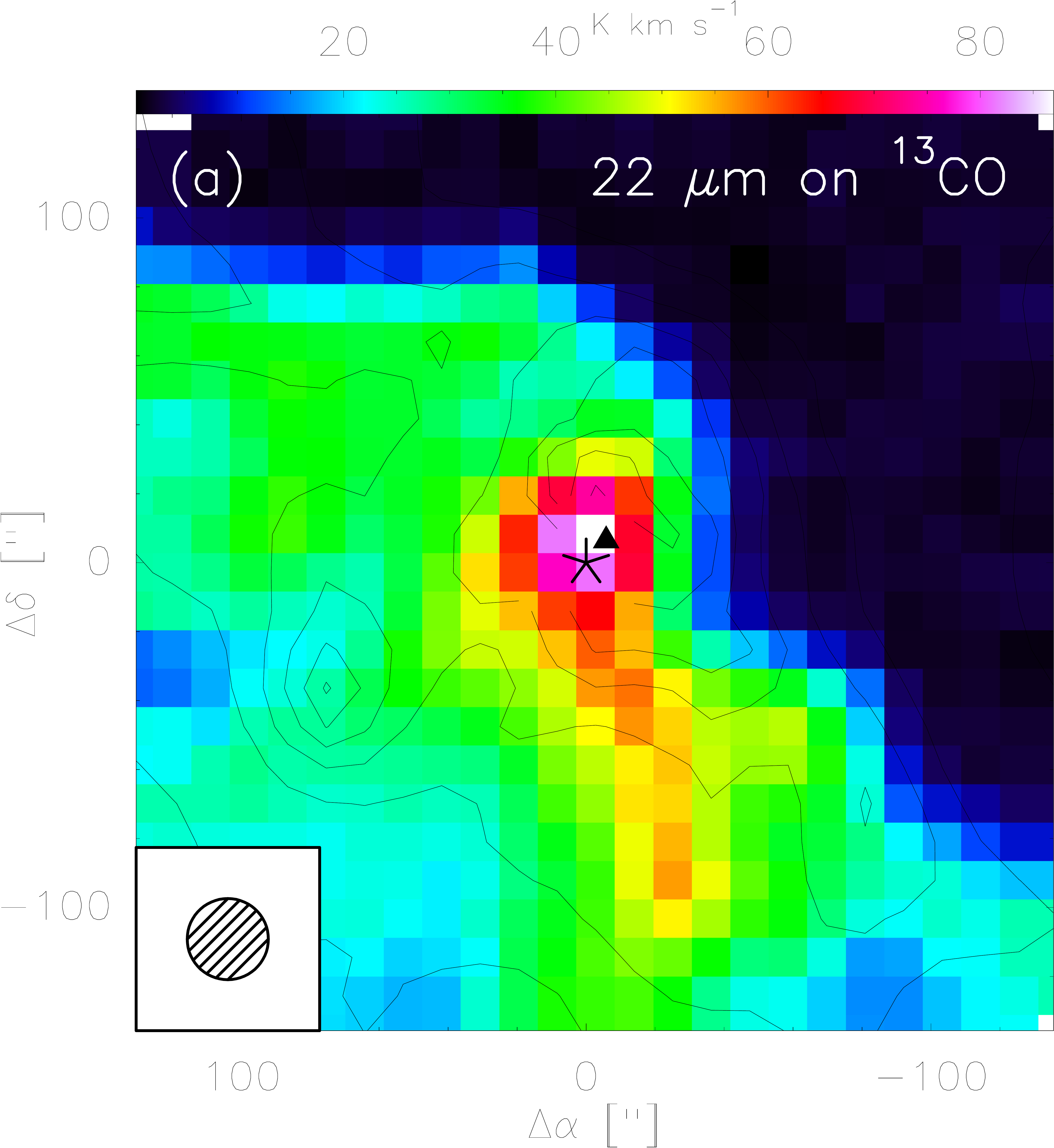}}\quad\quad
\subfigure{\includegraphics[width=80mm]{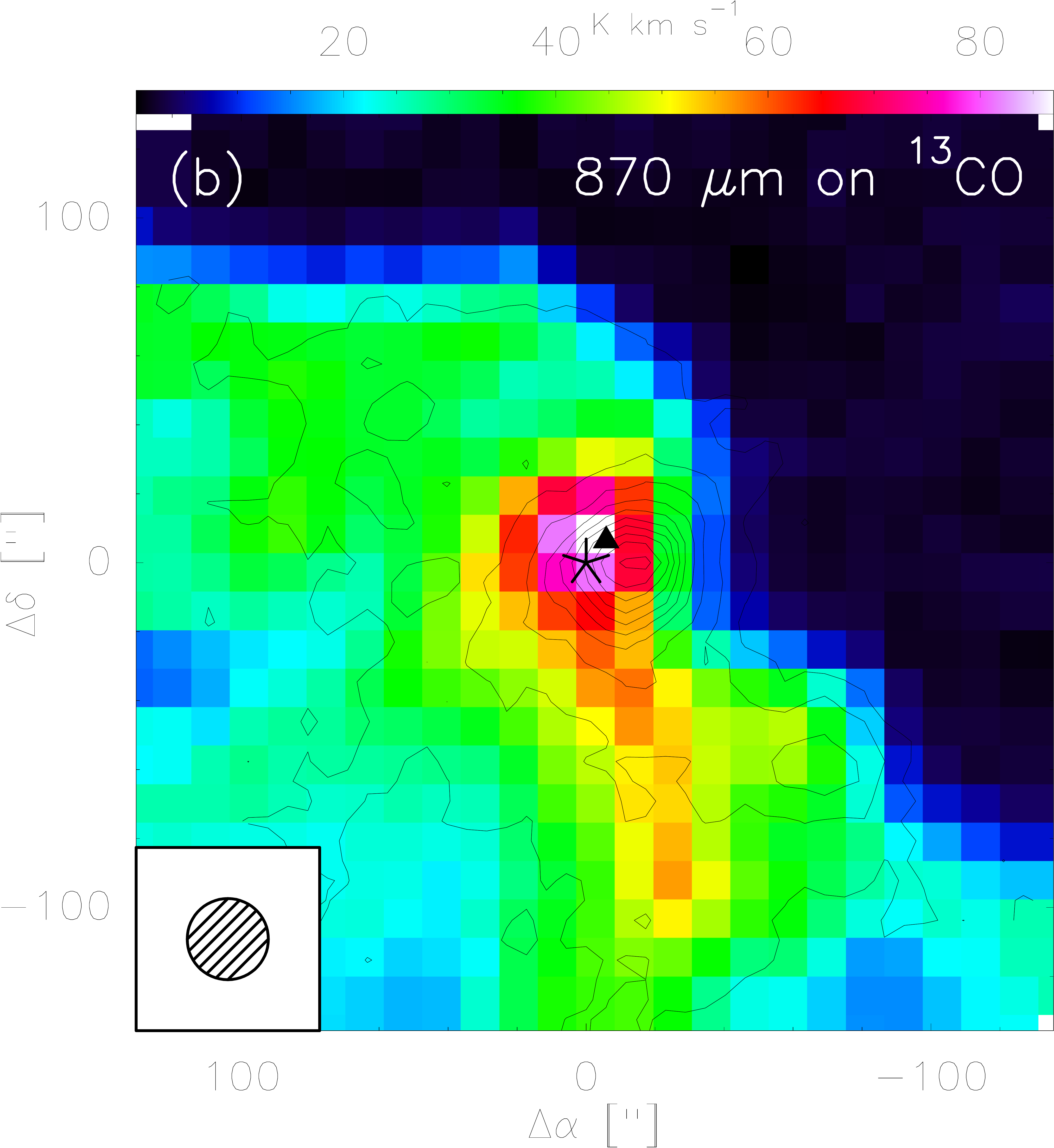}}
\subfigure{\includegraphics[width=80mm]{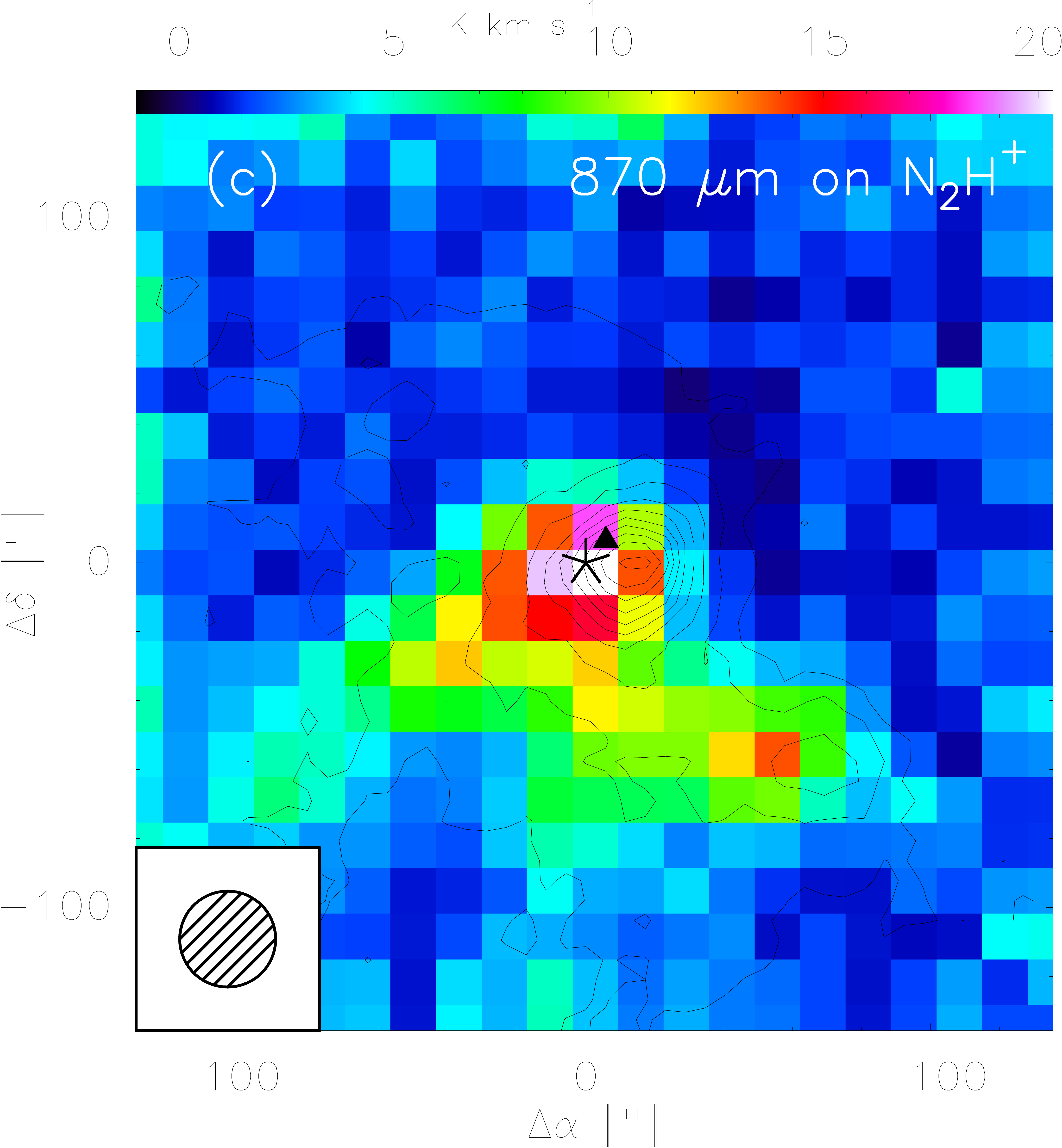}}\quad\quad
\subfigure{\includegraphics[width=80mm]{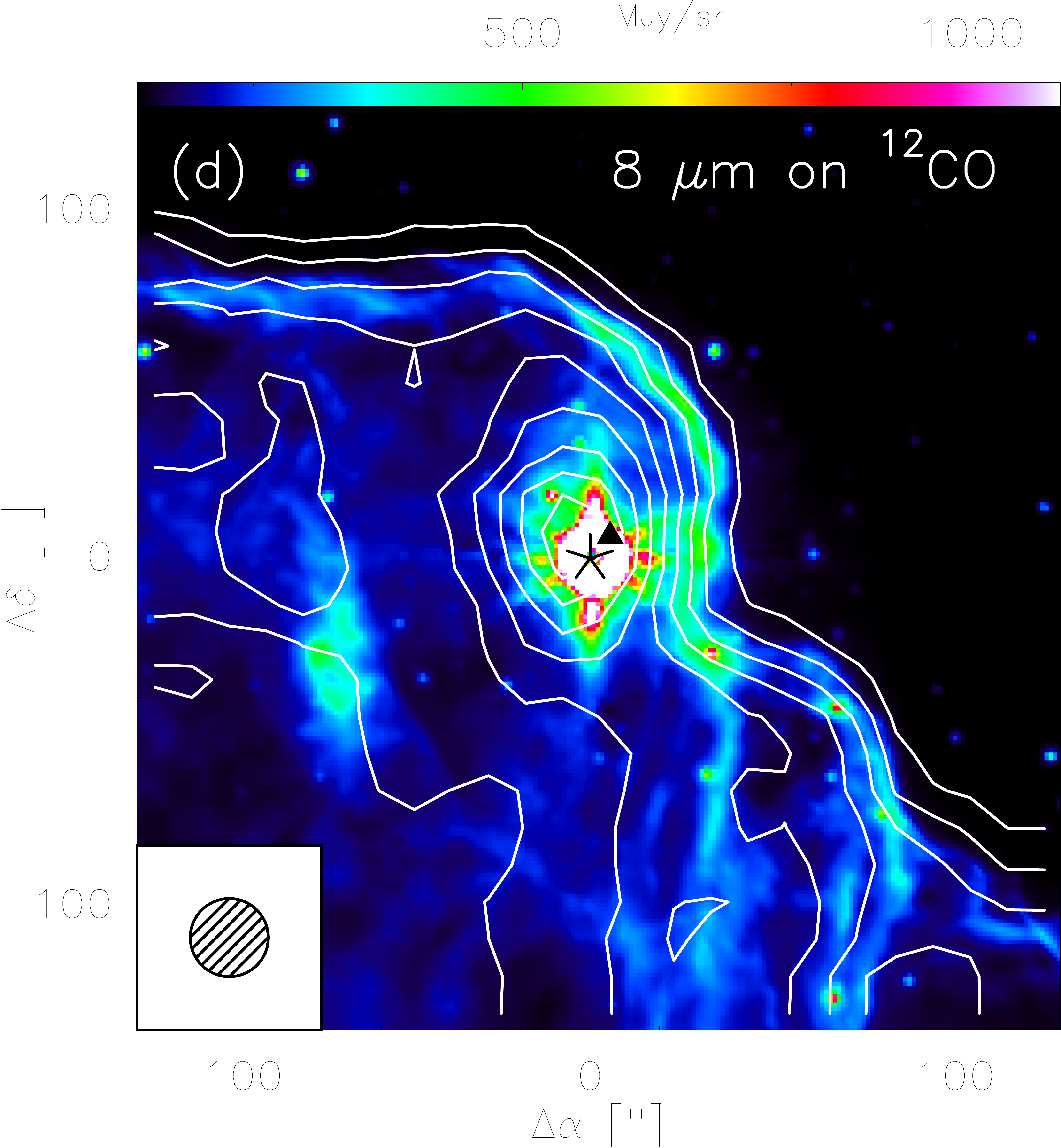}}

\caption{Top row: Color map of the velocity integrated intensity of $J$ = 1 $\to$ 0 transition of $^{13}$CO overlaid with contours of: (a), WISE 22~$\mu$m mid-IR continuum emission and (b), ATLASGAL 870~$\mu$m continuum emission toward M8~E. Bottom row: (c), shows the color map of the velocity integrated intensity of $J$ = 1 $\to$ 0 transition of N$_2$H$^+$ overlaid with ATLASGAL 870~$\mu$m dust continuum emission and (d), shows the contours of the velocity integrated intensity map of $^{12}$CO overlaid on the 8~$\mu$m PAH emission map. Position offsets are relative to the position of M8E-IR (marked with an asterisk) given in Section~2 and the ultracompact \hii\ region, M8E-radio, at ($\alpha$,$\delta$)$_{\rm J2000}$ = 18$^{\rm h}$04$^{\rm m}$52$^{\rm s}$8, -24$^\circ$26$\arcmin$36$\arcsec$ is represented by a black triangle. The $^{12}$CO and $^{13}$CO  emission maps have their original resolutions of 22.5$\arcsec$ and 23.5$\arcsec$, respectively,the MIPSGAL 24~$\mu$m continuum emission has a resolution of 6$\arcsec$, the ATLASGAL 870~$\mu$m continuum emission has a resolution of 18.2$\arcsec$ and the GLIMPSE 8~$\mu$m continuum emission has a 
resolution of 0.6$\arcsec$. The contour levels are 5\% to 100\% in steps of 10\% for the MIPSGAL~24~$\mu$m (a) and the ATLASGAL~870~$\mu$m (b, c) overlays. For the GLIMPSE~8~$\mu$m (d) overlay, the contour levels are 10\% to 100\% in steps of 10\%.} 
  
\label{ancillary_data} 
\end{figure*}

To explore the relationship between the cool molecular cloud and the hot ionized gas we compare our molecular line data with data from studies 
conducted at other wavelengths, mainly with the results of observations made in 
the IR and submm regimes. First, we extracted the 22~$\mu$m dust continuum image from the all-sky Wide-Field Infrared Survey Explorer (WISE, \citealt{2010AJ....140.1868W}). Mid-IR dust continuum emission probes the warm dust in \hii\ regions, which absorbs UV and far-UV radiation from a nearby massive star and re-emits in the MIR regime. The 22~$\mu$m emission wavelength is similar to that of the Multiband Infrared Photometer for Spitzer aboard the Spitzer Space Telescope (MIPSGAL, \citealt{2009PASP..121...76C}) whose 24~$\mu$m band data has been widely studied and found to be a direct tracer of \hii\ regions as which correlates very well with the 21~cm radio continuum emission, because the hot plasma that gives rise to the free-free thermal emission at radio wavelengths is ionised by the same sources that are responsible for the IR emission (\citealt{2010ApJ...718L.106B} and \citealt{2014ApJS..212....1A}). Second, we used the 870~$\mu$m dust continuum data from the APEX Telescope Large Area Survey of the Galaxy (ATLASGAL; \citealt{2009A&A...504..415S}), which is an unbiased survey covering 420 square degrees of the inner Galactic plane and was performed with the Large APEX BOlometer CAmera (LABOCA) instrument of the APEX 12~m telescope. Dust continuum emission in the submillimeter range probes dense and cool clumps in the ISM, hence tracing the early phases of (massive) star formation, including the earliest pre-stellar state. Figures~\ref{ancillary_data}~(a) and (b), show contours representing the 22~$\mu$m and the 870~$\mu$m emission overlaid on the velocity integrated intensity map of $J$ = 1 $\to$ 0 transition of $^{13}$CO represented by the color scale in the background. Both the IR and submm continuum emission peak very close to M8E-IR (marked with an asterisk) and M8E-radio (marked with a triangle). While the 24~$\mu$m emission is probing the warm dust heated by the \hii\ region, the 870~$\mu$m ATLASGAL emission follows the structure of the $^{13}$CO emission distribution better, with a bright extension toward the south-west of M8E-IR. To investigate further the nature of this bright south-west extension, we overlaid the contours of the ATLASGAL~870~$\mu$m emission on the velocity integrated intensity map of the N$_2$H$^+$  $J$ = 1 $\to$ 0  line (as the color scale background). As can be seen from Fig.~\ref{ancillary_data}~(c), The ATLASGAL contours 
follow the N$_2$H$^+$ emission distribution, including the secondary emission peak at N$_2$H$^{+}$ at ($\Delta\alpha$, $\Delta\delta$) = (--55$\arcsec$, --55$\arcsec$), hence probing the high density gas. We ascribe the slight offset ($\approx10''$) to the combined position uncertainties of the APEX and our IRAM 30 m telescope observations. Lastly, to explore the PDR environment, we used the 8~$\mu$m data from the archive of the Galactic Legacy Infrared Mid-Plane Survey Extraordinaire (GLIMPSE, \citealt{2003PASP..115..953B, Churchwell2009}) conducted with the Spitzer Space Telescope. The 8~$\mu$m band of the Infrared Array Camera (IRAC) used by Spitzer traces emission from the Polycyclic Aromatic Hydrocarbons (PAHs), large hydrocarbon molecules (or small dust grains) that get excited by strong UV radiation. The fluorescent IR emission from these PAHs, the result of FUV pumping,  arises from the PDR surface of dense molecular clouds and hence traces recent massive star formation activity \citep{2008tielens}. Figure~\ref{ancillary_data}~(d), shows the 8~$\mu$m PAH continuum emission in color scale overlaid with contours of the $^{12}$CO gas emission distribution. It can be seen that the high resolution 8~$\mu$m emission allows us to probe the bright rimmed structure enclosing the molecular cloud of M8~E. We believe that the central \hii\ region of M8 is expanding and this bright rimmed structure corresponds to its ionization front (IF) (discussed further in Section~5.2).

\section{Analysis}
We used several complementary methods to determine temperature and density of the gas responsible for the emission of various species observed toward M8~E with the IRAM~30~m telescope. We start with calculating the excitation temperatures and column densities of $^{12}$CO and $^{13}$CO using their $J$ = 1 $\to$ 0 transitions in Section~4.1. In Section~4.2, we used multi-line data of the dense cloud thermometer  CH$_3$CCH to determine the temperature of the dense gas, which then can be  used to determine the column densities of N$_2$H$^+$, HCN, H$^{13}$CN, HCO$^{+}$, H$^{13}$CO$^{+}$, HNC and HN$^{13}$C in Section~4.3. To complete our investigation of the physical conditions, we undertook non-LTE radiative transfer modeling to constrain the volume densities of the warm and cooler gas of M8~E in Section~4.4.       


\subsection{Excitation temperature and column density distributions of $^{12}$CO and $^{13}$CO}
When considering the radiative transfer of the line emission, we assumed that scattering is ignored and that the medium through which the radiation is traveling is uniform and described by a source function given by the Planck function with the level populations determined by an excitation temperature, $T_{\rm ex}$. 
Assuming the Rayleigh-Jeans approximation and taking the beam filling factor to be unity, we can express the observed main-beam brightness temperature, $T_{\rm mb}$, in terms of $T_{\rm ex}$ \citep[Eq.~1 of][]{2012A&A...538A..12P}. A detailed description of radiative transfer relevant here can be found in \citet{2015PASP..127..266M}. Assuming equal $T_{\rm ex}$ for $^{12}$CO and $^{13}$CO and that the $^{12}$CO emission is optically thick, the 
$T_{\rm ex}$ for the $J$ = 1 $\to$ 0 transition of $^{12}$CO can be defined as

\begin{equation}
     T_{\rm mb} = 5.53 \Bigg[\frac{1}{exp\frac{5.53}{T_{\rm ex}}-1} - \frac{1}{exp\frac{5.53}{T_{\rm bg}}-1} \Bigg] \rm K.
 \end{equation}

\begin{figure}[htp]
\subfigure{\includegraphics[width=80mm]{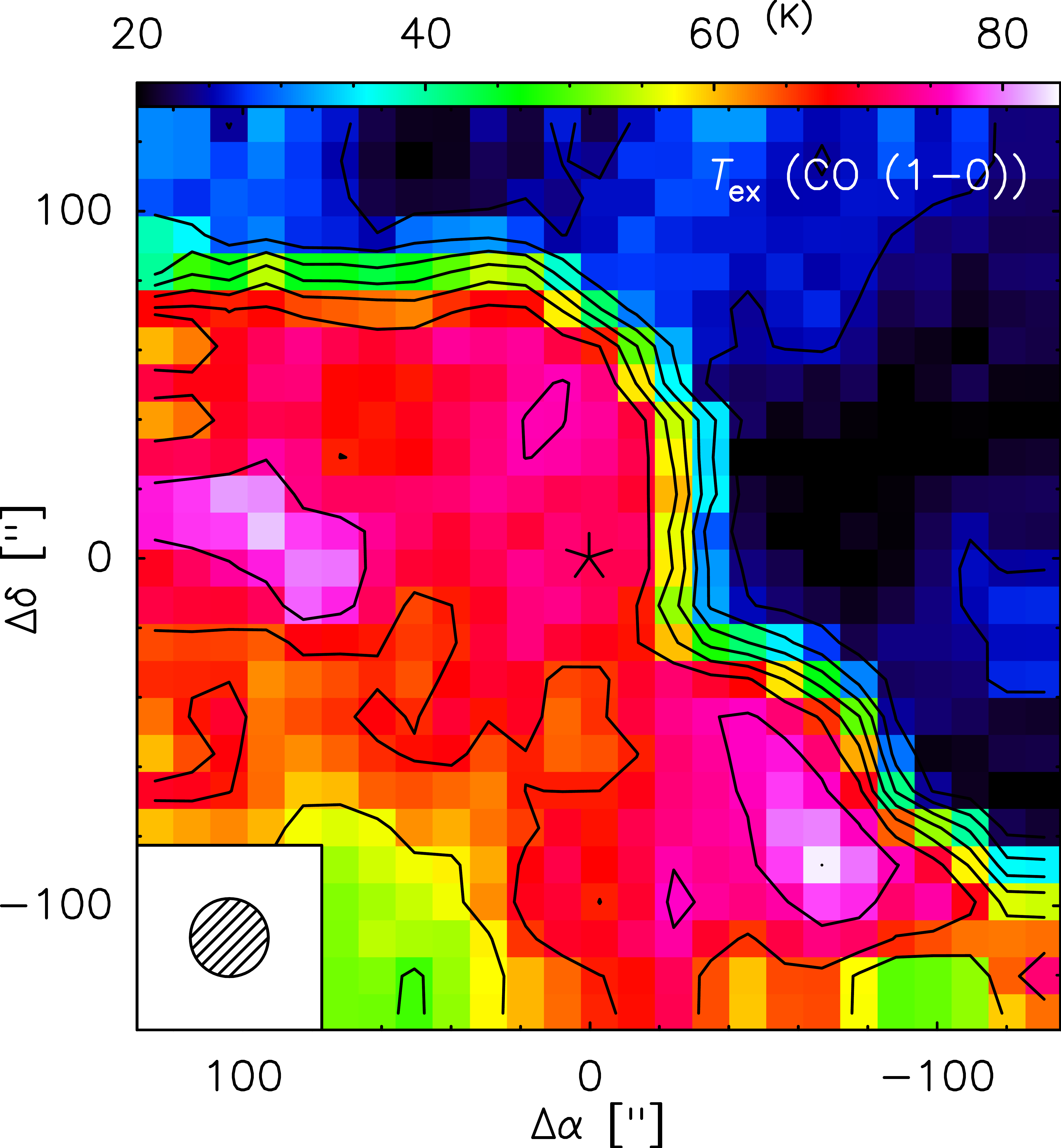}}
\subfigure{\includegraphics[width=80mm]{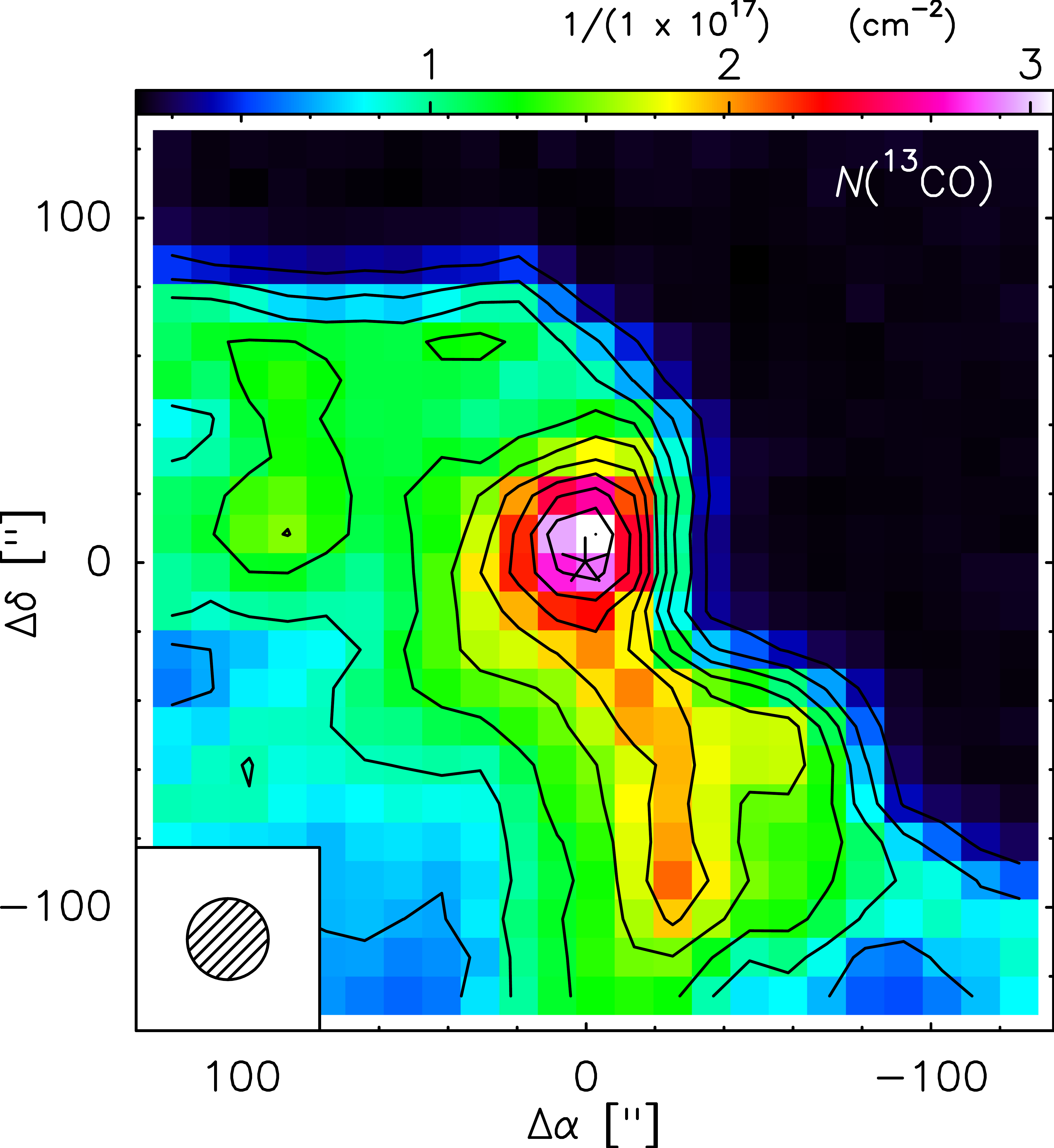}}

\caption{Upper panel: excitation temperature $T_{\rm ex}$ determined for the $J$ = 1 $\to$ 0 transition of $^{12}$CO and $^{13}$CO. Lower panel: total column density of $^{13}$CO, $N(\rm ^{13}CO)$. Position offsets are relative to the position of M8E-IR (marked with an asterisk) given in Section~2. The contour levels are from 10\% to 100\% in steps of 10\% peak emission (Table~\ref{all_obs}). The values of $T_{\rm mb}$ and that of $\int T_{\rm mb}(^{13}\rm CO)dv$ for the $J$ = 1 $\to$ 0 transition used to calculate $T_{\rm ex}$ and $N(\rm ^{13}CO)$ were extracted from maps convolved to the same resolution of 24$\arcsec$.} 
  
\label{tex_colden} 
\end{figure}

We took the background temperature, $T_{\rm bg}$ to be equal to that of the cosmic background radiation, i.e., 2.73~K. However, the contribution of the second term within the brackets in equation 1 will be $\le$~2\% of $T_{\rm mb}$ and hence can be neglected. 
The main-beam brightness temperature, $T_{\rm mb}$, is in K and its values are obtained from the peak $T_{\rm mb}$  map of $^{12}$CO in the velocity range of $\sim$ 0--20~km~s$^{-1}$. Figure~\ref{tex_colden}~upper panel shows the distribution of the resulting $T_{\rm ex}$ around M8E-IR, peaking in the east and south west of M8E-IR, with a maximum value of $\sim$ 85 $\pm$ 1~K. \

Under the assumption that the emission from $^{13}$CO is optically thin, we used the computed $T_{\rm ex}$ and the velocity integrated intensity, $\int T_{\rm mb}$($^{13}$CO) dv, to calculate the total column density of $^{13}$CO, $N(\rm ^{13}CO)$, by

\begin{equation}
     N(\rm ^{13}CO) = 4.5 \times 10^{13} (\textit{T}_{\rm ex} + 0.88)~exp\Bigg(\frac{5.3}{\textit{T}_{\rm ex}}\Bigg)~\int T_{\rm mb}(^{13}CO)~dv~cm^{-2}\,
 \end{equation}

\noindent where, $T_{\rm ex}$ is in K and $\int T_{\rm mb}(\rm ^{13}CO)$ dv is in K~km~s$^{-1}$. The resulting $N(\rm ^{13}CO)$ distribution is shown in Fig.~\ref{tex_colden}~(lower panel), peaking at M8E-IR with a maximum value of $\sim 3.0 \times 10^{17}$~cm$^{-2}$.
We calculated the H$_{2}$ column density, $N(\rm H_{2})$, with a peak value of $\sim$ 1.5 $\times$ 10$^{23}$~cm~$^{-2}$, by adopting an isotopic ratio of [CO/$^{13}$CO] $\sim$ 45, appropriate  for M8's Galactocentric distance \citep{2005ApJ...634.1126M} and a $^{12}$CO abundance ratio of [CO/H$_2$] $\sim$ 8.5 $\times$ 10$^{-5}$ \citep{2010pcim.book.....T}.

\subsection{CH$_3$CCH dense gas thermometry}
Methyl acetylene, CH$_3$CCH, was first detected by \cite{1973NPhS..243...45S} in Sgr B2. It is a symmetric top molecule with quantum numbers, $J$ and $K$, which define the total angular momentum and its projection along the symmetry axis, respectively.
For each energy level with total angular momentum $J$ 
there are is a ``ladder'' of  $J$ different $K$ states with  quantum numbers  $0 < K < J-1$. The states in each $K$ ladder span a wide range in energies above the ground, while the resulting lines 
have frequencies very close to each other. The electric dipole moment of CH$_3$CCH is parallel to the symmetry axis, so the only possible $J$ $\to$ $J - 1$ transitions have $\Delta$ $K$ = 0. For a given $J$ level, the different $K$ levels are populated only through collisions and hence the total population in each $K$ ladder is a function of only the kinetic temperature of the dense gas emitting the  CH$_3$CCH lines. The temperature can be determined from the (measured) relative level populations. This makes CH$_3$CCH an excellent molecular cloud thermometer (e.g. \citealt{1994ApJ...431..674B}, \citealt{2012A&A...538A..41G} and \citealt{2017A&A...603A..33G}). 

To derive the dense gas temperature, we simultaneously fitted the $J$ = 5 $\to$ 4 and 6 $\to$ 5 transitions of CH$_3$CCH observed toward M8E-IR. We used MCWeeds \citep{2017A&A...603A..33G}, which provides an external interface between Weeds (a CLASS extension for the analysis of multi molecule/multi transition millimeter and submillimeter spectral surveys, \citealt{2011A&A...526A..47M}) and PyMC\footnote{https://pymc-devs.github.io/pymc/index.html} (a python package to efficiently code a probabilistic model and draw samples from its posterior distribution using Markov chain Monte Carlo techniques, \citealt{JSSv035i04}). Weeds generates synthetic spectra by solving the radiative transfer equation assuming LTE and takes into account the finite angular resolution of the observations, but does not optimise algorithms. On the other hand, PyMC implements Bayesian statistical models and fit algorithms. Initial guesses are given for a set of input parameters along with their probability distribution and the range over which they should be varied. By considering multiple transitions of a molecule, McWeeds provides best fit parameters with their errors (for more details see \citealt{2017A&A...603A..33G} and \citealt{2019A&A...623A..68T}). 

\begin{table}
\label{chch}
\tiny
\centering
\begin{threeparttable}
\caption{Observed spectral line parameters of CH$_3$CCH.}
\renewcommand{\arraystretch}{1.4}
\begin{tabular}{c c c c c}
\hline\hline
  Transition & Frequency (GHz) & $E_{\rm up}$ (K) & $T_{\rm peak}$ (K) & rms (K) \\
  \hline
 $J$, $K$ = 5, 4 $\to$ 4, 4 & 85.4312 & 127.9 & $<$ 3 $\times$ rms & \multirow{5}{*}{0.026} \\
 $J$, $K$ = 5, 3 $\to$ 4, 3 & 85.4425 & 77.3 & 0.137 &  \\
 $J$, $K$ = 5, 2 $\to$ 4, 2 & 85.4507 & 41.2 & 0.204 &  \\
 $J$, $K$ = 5, 1 $\to$ 4, 1 & 85.4556 & 19.5 & 0.464 &  \\
 $J$, $K$ = 5, 0 $\to$ 4, 0 & 85.4572 & 12.3 & 0.630 &  \\
\hline 
 $J$, $K$ = 6, 5 $\to$ 5, 5 & 102.4991 & 197.8 & $<$ 3 $\times$ rms  & \multirow{6}{*}{0.030} \\
 $J$, $K$ = 6, 4 $\to$ 5, 4 & 102.5165 & 132.8 & $<$ 3 $\times$ rms  &  \\
 $J$, $K$ = 6, 3 $\to$ 5, 3 & 102.5303 & 82.3 & 0.162 &  \\
 $J$, $K$ = 6, 2 $\to$ 5, 2 & 102.5401 & 46.1 & 0.298 &  \\
 $J$, $K$ = 6, 1 $\to$ 5, 1 & 102.5460 & 24.5 & 0.696 &  \\
 $J$, $K$ = 6, 0 $\to$ 5, 0 & 102.5479 & 17.2 & 0.796 &  \\
 
\hline
\end{tabular}


 \label{ch3cch_table}
 \end{threeparttable}
\end{table}

To fit the spectrum, an input file is provided containing all the initial guesses of the parameters: column density (cm$^{-2}$), temperature (K), size of the emitting region (arcsec), radial velocity (km~s$^{-1}$) and line width (km~s$^{-1}$), rms noise (K), and calibration uncertainty for the spectral range. The input file also contains the name of the species, the frequency range in which the interested transition lies, the corresponding data file and the priors required to fit the observed spectrum. Using the ATLASGAL~870~$\mu$m image, we estimated the source size $\sim$ 40$\arcsec$, which is larger than the beam size ($\sim$ 30$\arcsec$) with which CH$_3$CCH lines were observed. Therefore, we fix the size of the emitting region to 30$\arcsec$. Similar to \citet{2017A&A...603A..33G}, we used loosely informative priors for the parameters of the fit: A normal Gaussian function for column densities and velocities, while a truncated normal Gaussian function is used for temperatures and line widths. The probability distributions of these functions depend on the values of their mean, $\mu$, and standard deviation, $\sigma$ (A detailed description of priors can be found in Jaynes \& Bretthorst (2003)).

\begin{figure*}[htp]
\centering
\subfigure{\includegraphics[width=82mm]{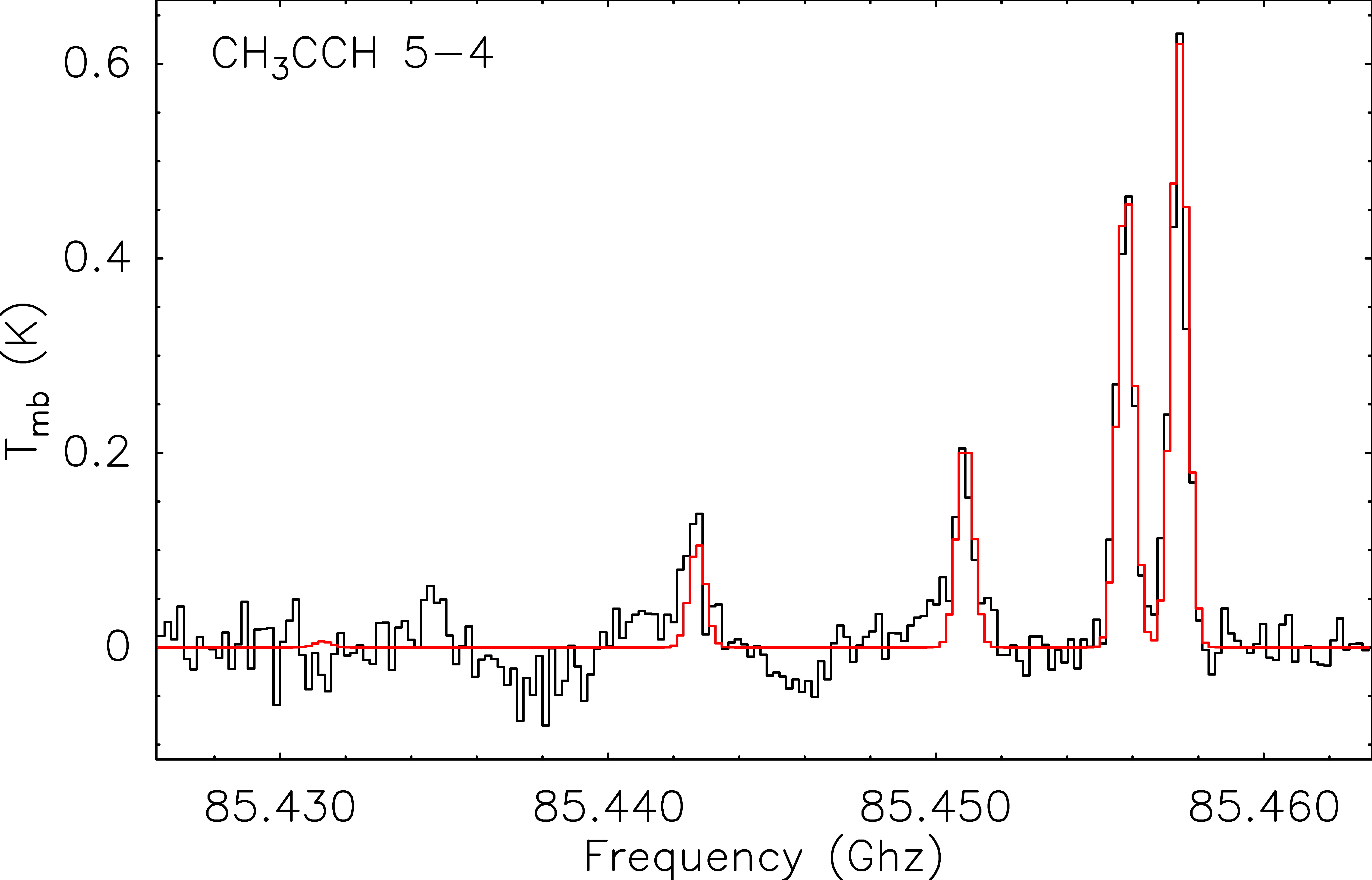}}\quad\quad
\subfigure{\includegraphics[width=82mm]{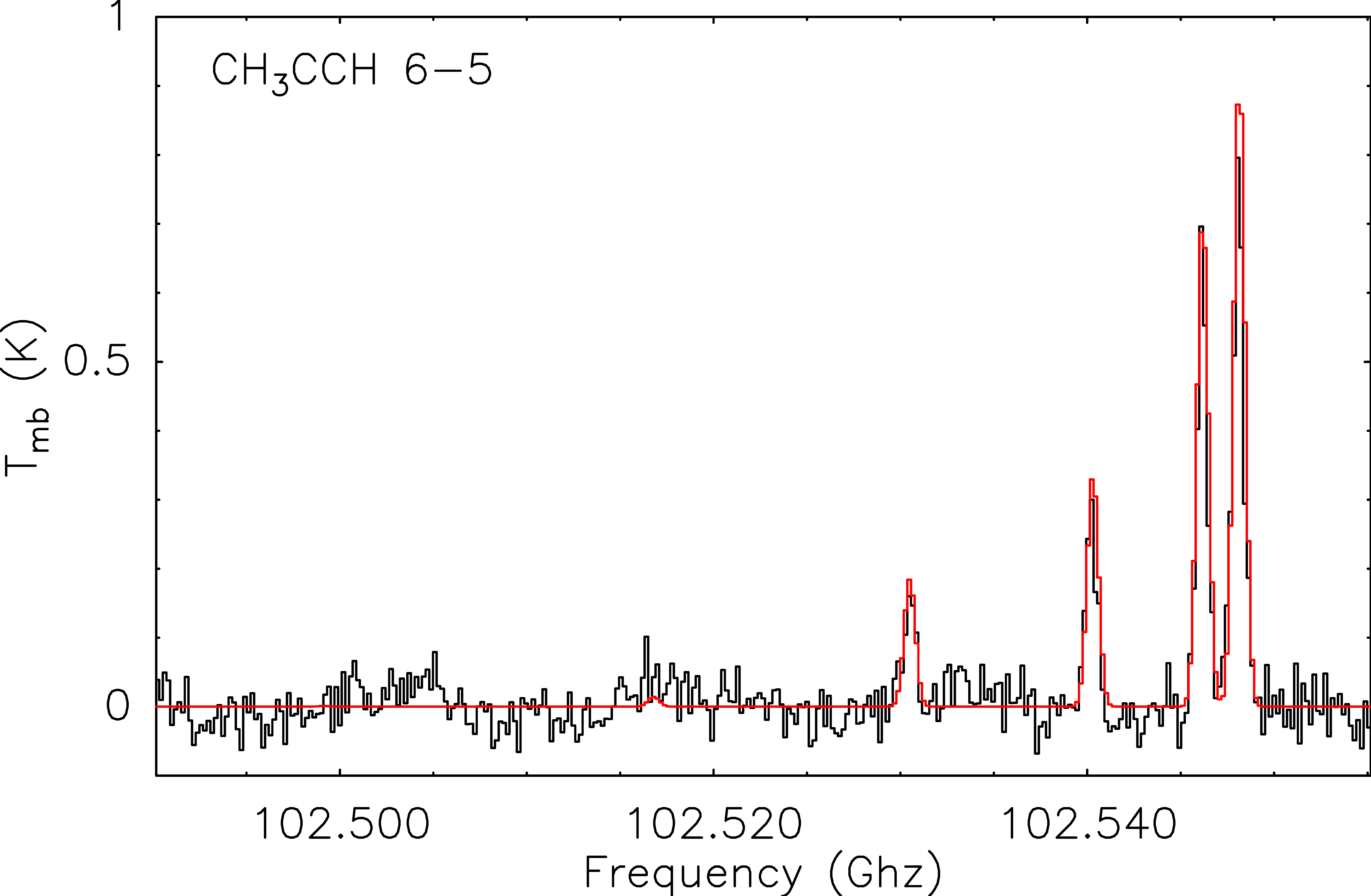}}

\caption{Spectra of the $J$ = 5 $\to$ 4 and 6 $\to$ 5 transitions of CH$_3$CCH observed toward M8E-IR. In red are the best fit synthetic spectra generated by MCWeeds for the input parameters listed in Table~\ref{mcweeds}.} 
  
\label{ch3cch} 
\end{figure*}

The given input (initial guesses along with their upper and lower bounds) and the obtained output values of the fit parameters used in MCWeeds are listed in Table~\ref{mcweeds}. The resulting dense gas temperature comes out as $32.9 \pm 1.4$~K, which is in excellent agreement with the value for the kinetic temperature of the molecular gas toward M8~E (31~K) determined from measurements of NH$_3$ emission with a comparable beam size ($40''$ FWHM) \citep{1996A&A...308..573M}. A total CH$_3$CCH column density, $N$(CH$_3$CCH), of $(5.3 \pm 0.1) \times 10^{14}$~cm$^{-2}$ was obtained.



\subsection{Column density estimates of N$_2$H$^+$, HCN, HCO$^{+}$, HNC, H$^{13}$CN, H$^{13}$CO$^{+}$ and HN$^{13}$C} 
In Sections~4.1 and 4.2, we estimated the temperatures of the warm, low density and the cooler, higher density gas components based on $^{12}$CO and CH$_3$CCH emission, respectively. We assume that the temperature of the cool gas ($\sim$ 32~K) estimated from CH$_3$CCH is similar to the kinetic temperature of the gas responsible for the emission of N$_2$H$^+$, HCN, HCO$^{+}$, HNC, H$^{13}$CN, H$^{13}$CO$^{+}$ and HN$^{13}$C. Thus we calculated these molecules' total column densities by fitting the synthetic spectra generated by MCWeeds to the observed spectra of their $J$ = 1 $\to$ 0 transition toward M8E-IR, keeping temperature and size fixed. The input and output parameters are listed in Table~\ref{mcweeds}. The calculated column densities of HCN, HCO$^{+}$ and HNC should be considered as upper limits to the real values as we assumed the cool gas temperature for their evaluation. We compared our results with a similar survey done toward W~51, where \citet{2017ApJ...845..116W} assume lower excitation temperatures ($\le$ 20~K) and we find that the total column densities determined for the observed species in M8~E are relatively higher.




 \begin{table*}[htbp]
\small
\centering
\begin{threeparttable}
\caption{Priors, given input and resulting output from the best-fit models.}
\begin{tabular}{c c c c c}
\hline\hline
\noalign{\smallskip}
  & Temperature (K) & Column density (Log(cm$^{-2}$)) & line width (km~s$^{-1}$) & velocity (km~s$^{-1}$) \\
  \hline
 \noalign{\smallskip}
 \multicolumn{5}{c}{CH$_3$CCH $J$ = 5 $\to$ 4 \& 6 $\to$ 5}\\
 \hline
 \noalign{\smallskip}
 Prior & Truncated normal & Normal &Truncated normal & Normal\\
 \multirow{3}{*}{Input} & $\mu$ = 50  & $\mu$ = 14  & $\mu$ = 5 & $\mu$ = 0\\
 & $\sigma$ = 30 & $\sigma$ = 2 & $\sigma$ = 3 & $\sigma$ = 2 \\
 & low = 10, high = 80 & & low = 1, high = 35 & \\
 \hline
 \noalign{\smallskip}
 Output & 32.9 (1.4) & 14.72 (0.047) & 2.03 (0.05) & -0.24 (0.019) \\
 \hline
 \noalign{\smallskip}
 \multicolumn{5}{c}{N$_2$H$^+$ $J$ = 1 $\to$ 0}\\
 \hline
 \noalign{\smallskip}
 Prior & Fixed & Normal &Truncated normal & Normal\\
 \multirow{3}{*}{Input} & \multirow{3}{*}{value = 33}  & $\mu$ = 14  & $\mu$ = 5 & $\mu$ = 0\\
 &  & $\sigma$ = 2 & $\sigma$ = 3 & $\sigma$ = 2 \\
 &  & & low = 1, high = 35 & \\
 \hline
 \noalign{\smallskip}
 Output & 33 & 14.11 (0.042) & 2.24 (0.03) & -0.28 (0.01) \\
 \hline
 \noalign{\smallskip}
 \multicolumn{5}{c}{HCN $J$ = 1 $\to$ 0}\\
 \hline
 \noalign{\smallskip}
 Prior & Fixed & Normal &Truncated normal & Normal\\
 \multirow{3}{*}{Input} & \multirow{3}{*}{value = 33}  & $\mu$ = 14  & $\mu$ = 5 & $\mu$ = 0\\
 &  & $\sigma$ = 2 & $\sigma$ = 3 & $\sigma$ = 2 \\
 &  & & low = 1, high = 35 & \\
 \hline
 \noalign{\smallskip}
 Output & 33 & 15.24 (0.042) & 3.07 (0.08) & -0.37 (0.03) \\
 \hline
 \noalign{\smallskip}
 \multicolumn{5}{c}{H$^{13}$CN $J$ = 1 $\to$ 0}\\
 \hline
 \noalign{\smallskip}
 Prior & Fixed & Normal &Truncated normal & Normal\\
 \multirow{3}{*}{Input} & \multirow{3}{*}{value = 33}  & $\mu$ = 14  & $\mu$ = 5 & $\mu$ = 0\\
 &  & $\sigma$ = 2 & $\sigma$ = 3 & $\sigma$ = 2 \\
 &  & & low = 1, high = 35 & \\
 \hline
 \noalign{\smallskip}
 Output & 33 & 13.84 (0.037) & 2.25 (0.02) & -0.22 (0.008) \\
 \hline
 \noalign{\smallskip}
 \multicolumn{5}{c}{HCO$^+$ $J$ = 1 $\to$ 0}\\
 \hline
 \noalign{\smallskip}
 Prior & Fixed & Normal &Truncated normal & Normal\\
 \multirow{3}{*}{Input} & \multirow{3}{*}{value = 33}  & $\mu$ = 14  & $\mu$ = 5 & $\mu$ = 0\\
 &  & $\sigma$ = 2 & $\sigma$ = 3 & $\sigma$ = 2 \\
 &  & & low = 1, high = 35 & \\
 \hline
 \noalign{\smallskip}
 Output & 33 & 14.84 (0.027) & 2.45 (0.038)  & -0.13 (0.009) \\
 \hline
 \noalign{\smallskip}
 \multicolumn{5}{c}{H$^{13}$CO$^+$ $J$ = 1 $\to$ 0}\\
 \hline
 \noalign{\smallskip}
 Prior & Fixed & Normal &Truncated normal & Normal\\
 \multirow{3}{*}{Input} & \multirow{3}{*}{value = 33} & $\mu$ = 14  & $\mu$ = 5 & $\mu$ = 0\\
 &  & $\sigma$ = 2 & $\sigma$ = 3 & $\sigma$ = 2 \\
 &  & & low = 1, high = 35 & \\
 \hline
 \noalign{\smallskip}
 Output & 33 & 12.97 (0.045) & 2.1 (0.03) & -0.23 (0.01) \\
 \hline
 \noalign{\smallskip}
 \multicolumn{5}{c}{HNC $J$ = 1 $\to$ 0}\\
 \hline
 \noalign{\smallskip}
 Prior & Fixed & Normal &Truncated normal & Normal\\
 \multirow{3}{*}{Input} & \multirow{3}{*}{value = 33}  & $\mu$ = 14  & $\mu$ = 5 & $\mu$ = 0\\
 &  & $\sigma$ = 2 & $\sigma$ = 3 & $\sigma$ = 2 \\
 &  & & low = 1, high = 35 & \\
 \hline
 \noalign{\smallskip}
 Output & 33 & 14.35 (0.05) & 2.59 (0.032) & -0.12 (0.007) \\
 \hline
  \noalign{\smallskip}
 \multicolumn{5}{c}{HN$^{13}$C $J$ = 1 $\to$ 0}\\
 \hline
 \noalign{\smallskip}
 Prior & Fixed & Normal &Truncated normal & Normal\\
 \multirow{3}{*}{Input} & \multirow{3}{*}{value = 33}  & $\mu$ = 14  & $\mu$ = 5 & $\mu$ = 0\\
 &  & $\sigma$ = 2 & $\sigma$ = 3 & $\sigma$ = 2 \\
 &  & & low = 1, high = 35 & \\
 \hline
 \noalign{\smallskip}
 Output & 33 & 13.24 (0.04) & 2.1 (0.043)  & -0.06 (0.01) \\
 \hline

\end{tabular}


 \label{mcweeds}
 \end{threeparttable}
\end{table*}

\subsection{Non-LTE analysis}
In Sections~4.1--4.3, we used several techniques under the assumption of LTE to determine the column densities of the observed species. To check how much these values deviate under non-LTE conditions, we used RADEX \citep{2007A&A...468..627V}, which is a non-LTE radiative transfer program.










\begin{table}
\tiny
\centering
\begin{threeparttable}
\caption{Comparison of column densities determined from RADEX and LTE analysis.}
\renewcommand{\arraystretch}{1.4}
\begin{tabular}{c |c c c|c}
\hline\hline
  Species & \multicolumn{3}{c|}{non-LTE (RADEX)} & LTE\\
  \hline
 & $T_{\rm k}$ (K) & $n(\rm H_2)$ (cm$^{-3}$) & $N$(X) (cm$^{-2}$) & $N$(X) (cm$^{-2}$)\\
  \hline
 $^{12}$CO & 80 & 10$^4$ & 7.8 $\times$ 10$^{18}$ & 1.3 $\times$ 10$^{19}$ \\
$^{13}$CO & 80 & 10$^4$ & 3.6 $\times$ 10$^{17}$ & 2 $\times$ 10$^{17}$\\
N$_2$H$^{+}$ & 30 & 3.7 10$^5$ & 1 $\times$ 10$^{14}$ & 5.2 $\times$ 10$^{14}$ \\ 
HCN & 30 & 1.5 $\times$ 10$^5$ & 1 $\times$ 10$^{15}$ & 1.7 $\times$ 10$^{15}$\\
HCO$^{+}$ & 30 & 3.2 $\times$ 10$^4$ & 5 $\times$ 10$^{14}$ & 7 $\times$ 10$^{14}$ \\
H$^{13}$CO$^{+}$ & 30 & 1.8 $\times$ 10$^5$ & 7 $\times$ 10$^{12}$ & 9.3 $\times$ 10$^{12}$ \\
HNC & 30 & 8 $\times$ 10$^4$ & 4.2 $\times$ 10$^{14}$ & 2.2 $\times$ 10$^{14}$ \\

\hline
\end{tabular}


 \label{radex}
 \end{threeparttable}
\end{table}

RADEX is a computer code for performing statistical equilibrium calculations that treats the radiative transfer using the escape probability approximation in an isothermal and homogeneous medium, taking into account optical depth effects. We selected a uniform spherical geometry and the collisional rate coefficients, required for modeling, were taken from the Leiden Molecular and Atomic Database\footnote{https://home.strw.leidenuniv.nl/~moldata/} (LAMDA; \citealt{2005A&A...432..369S}). With the exception of H$^{13}$CN, CH$_3$CCH and HN$^{13}$C, LAMDA provides collisional rate coefficients for all species analysed in this paper. For $^{12}$CO and $^{13}$CO, the collisional rate coefficients with H$_2$ are adopted from \citet{2010ApJ...718.1062Y}. The N$_2$H$^+$--H$_2$ collisional rate coefficient are calculated (see \citealt{2005A&A...432..369S}) from the N$_2$H$^+$--He collisional rate coefficients given by \citet{2005MNRAS.363.1083D} for hfs components. The HCN--H$_2$ collisional rate coefficients have been calculated by \citet{2017MNRAS.468.1084H} and the coefficients for the hfs components are computed following Braine et al. (2020, in prep). The H$^{13}$CO$^+$--H$_2$ collisional rate coefficients are extrapolated from the HCO$^+$--H$_2$ collisional rate coefficients, which are provided by \citet{1999MNRAS.305..651F}. The HNC--H$_2$ collisional rate coefficients are calculated by multiplying by a factor of 1.37 the HNC-He collision rates given by \citet{2010MNRAS.406.2488D}.

Since for the di- and triatomics we only have data  for a single transition each, volume densities cannot be constrained and we were guided by the concepts of critical, $n_{\rm cr}$, and effective, $n_{\rm eff}$, densities as estimates for the densities required to excite the observed lines. \citet{2015PASP..127..299S} thoroughly compares the relevance of $n_{\rm cr}$ and $n_{\rm eff}$. The common usage of critical density $n_{\rm cr}$ applies for the optically thin limit and ignores radiative trapping \citep{2015PASP..127..299S}. Since, in particular for ground state lines, the line opacities can be very high, it should only be considered as an upper limit for the density needed to excite the molecule's rotational levels \citep{2017A&A...599A..98P}. For optically thick lines, the effective density becomes relevant. Using the $^{12}$C/$^{13}$C $\sim$ 45 \citep{2005ApJ...634.1126M}, we calculated the optical depths $\tau_{\rm ^{12}CO}$ $\sim$ 19, $\tau_{\rm HCN}$ $\sim$ 7, $\tau_{\rm HCO^{+}}$ $\sim$ 6 and $\tau_{\rm HNC}$ = 3.9. For N$_2$H$^+$, we fitted its different hfs components using the CLASS software and obtained $\tau_{\rm N_2H^+}$ $\sim$ 0.6. For optically thick lines, $n_{\rm eff}$ = $n_{\rm cr}$ $\times$ (1 - exp(-$\tau$))/$\tau$ \citep{2015PASP..127..299S}.


Assuming a background temperature of 2.73~K and taking the line width values of the above mentioned species from the observed spectra (toward M8E-IR), we ran the models for fixed kinetic temperatures and volume densities, while varying the column densities to reach the intensities that best fit our observed values. Kinetic temperatures of 80~K and 30~K were used as derived above from $^{12}$CO and CH$_3$CCH, respectively. We set the volume densities (given in Table~\ref{radex}) same as $n_{\rm cr}$ for optically thin species ($^{13}$CO, H$^{13}$CO$^+$ and N$_2$H$^+$) and $n_{\rm eff}$ for optically thick species ($^{12}$CO, HCN, HCO$^{+}$ and HNC). 




As mentioned in Sect.~4.2, the LTE analysis was done for a fixed beam size of 30$''$ and the source size is $\sim$ 40$''$. Thus, to compare the RADEX modeling results with the LTE analysis, we divide the column density values predicted by RADEX by a resulting beam filling factor of 0.64 and these results are reported in Table~\ref{radex}. For $^{12}$CO, with a kinetic temperature of 80~K and a $n_{\rm eff}$ $\sim$ 10$^2$~cm$^{-3}$, the models do not reach the peak intensity of $\sim$ 72~K. We require volume densities $>$ 10$^4$~cm$^{-3}$ in order to reach the observed peak intensity. The corresponding column density is given in Table~\ref{radex}. For consistency, we use a temperature of 80~K and volume density of 10$^4$~cm$^{-3}$ for $^{13}$CO as well. If we use lower densities ($n_{\rm cr}$ = 2 $\times$ 10$^3$~cm$^{-3}$), we will get much higher column density values than estimated by the LTE analysis. In fact the column density has an inverse dependence on $n_{\rm eff}$ \citep{2015PASP..127..299S}. Assuming higher volume densities than $n_{\rm cr}$ and $n_{\rm eff}$ can also be justified, given that we detect lines from higher density probes such as N$_{2}$H$^+$ toward M8~E. For the rest of the species, we get similar column densities as those calculated from the LTE analysis.



\begin{figure*}[htp]
\centering
\subfigure{\includegraphics[width=185mm]{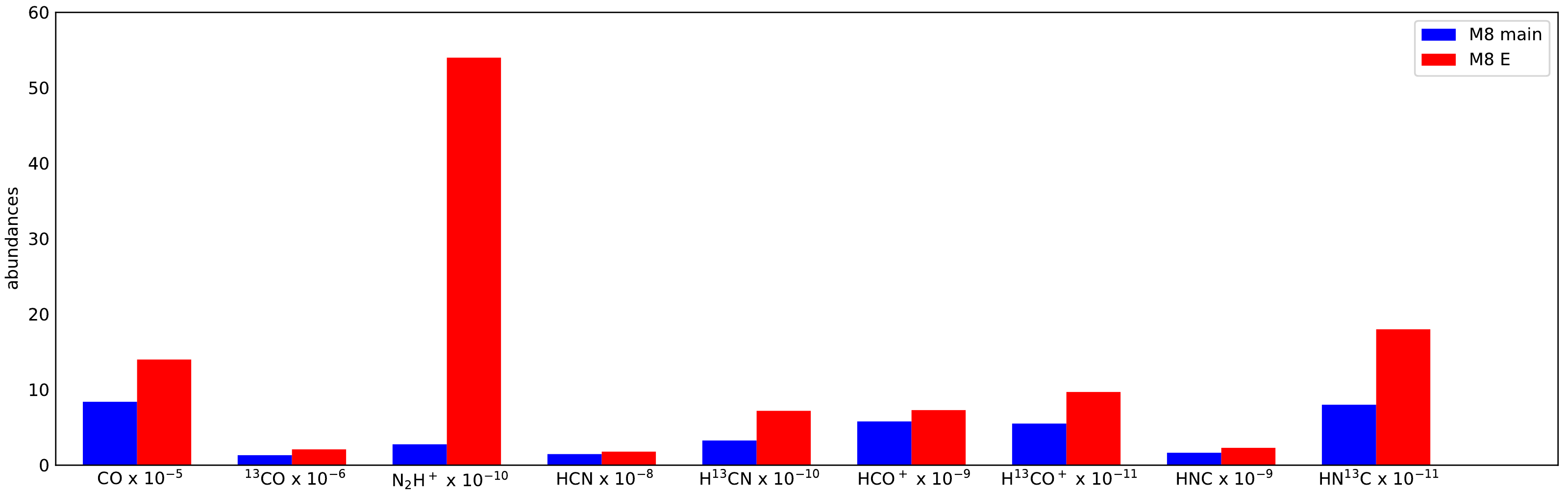}}

\caption{Chemical abundances of $^{12}$CO, $^{13}$CO, HCN, H$^{13}$CN, HCO$^{+}$, H$^{13}$CO$^{+}$, HNC and HN$^{13}$C observed toward M8~E (red bars) and M8-Main (blue bars).} 
  
\label{histogram} 
\end{figure*}




\section{Discussion}

\subsection{Comparison with M8-Main}
Given the proximity of M8~E to M8-Main and the fact that the two rival each other at IR and submillimeter wavelengths \citep{2008hsf2.book..533T}, it is only natural to compare the two regions based on their chemical abundances and physical conditions. A comparison of the abundances, $N$(X)/$N$(H$_2$), of various species observed toward M8-Main and M8~E is presented in Fig.~\ref{histogram}. For M8~E, the abundances are calculated from the column densities of various species, $N$(X), obtained from the LTE analysis done in Sections~4.1--4.3, while the H$_2$ column density is calculated using the intensity of 870~$\mu$m continuum emission measured in the ATLASGAL survey, which is 5588~mJy~beam$^{-1}$ toward M8E-IR, and by assuming the  dust temperature, $T_{\rm d}$, of 28.6~K determined by \citep{2002ApJ...580..285T}. We assumed a gas-to-dust mass ratio of 100 and an absorption coefficient of 1.85~g$^{-1}$~cm$^2$ \citep{2009A&A...504..415S}. Using eq.~A.27 from \citet{2008A&A...487..993K}, we calculate a peak molecular hydrogen column density,  $N$(H$_2$), of 1.46 $\times$ 10$^{22}$~cm$^{-2}$, which is in reasonable agreement with the values discussed in Section~4.1, within a factor $\sim$ 1.5. For M8-Main we computed the column densities toward Her~36, which is the brightest line-of-sight (similar to M8E-IR in M8~E). We adopted the $^{12}$CO and $^{13}$CO column densities as determined from the LTE analysis in \citet[Section~4.1]{refId0} and for the other molecules analysed in the work, we used the same techniques as mentioned in Sections~4.2 and 4.3. A full description of the 3~mm observations toward M8-Main is given in \citet[Section~2.2]{2019A&A...626A..28T}.

From Fig.~\ref{histogram}, it can be seen that in general the abundance of each species is found to be higher toward M8~E than in M8-Main. Though for most of the molecules the difference is small, for N$_2$H$^+$ and HN$^{13}$C, the abundances found toward M8~E are about 25 times and 2 times, respectively, larger than in M8-Main. Since the $J$ = 1 $\to$ 0 transition of N$_2$H$^+$ is argued to be a better tracer of very dense gas, which is not relatively bright in M8-Main, compared to the other traditional molecules like HCN and HCO$^+$ (e.g. \citealt{2017A&A...605L...5K}, \citealt{2017A&A...599A..98P} and \citealt{2020arXiv200306842B}), we infer that in M8~E, star formation is occurring in a densely embedded molecular cloud core that has no counterpart in M8-Main, which in general represents a warmer and more diffuse PDR dominated environment. 
This is consistent with obvious picture 
that the molecular cloud of M8~E, powered by M8E-IR, which will soon emerge as  BO type star, is at an earlier stage of star formation as M8-Main, in which more massive young stellar objects, represented by the Her~36 system and other O-type stars are already exciting a prominent \hii\ region.

\subsection{Spatial structure relative to M8's Ionization front}

The large-scale structure of the Lagoon Nebula has been discussed based on optical, radio and IR data by \cite{1976ApJ...203..159L}, \cite{1986AJ.....91..870W} \& \cite{2008hsf2.book..533T} and the zoomed in view of the M8-Main region 
in \cite{refId0}. Combining our previous knowledge of the Lagoon Nebula and the information we attained from analysing the observed spatial distribution of the molecular line emission toward M8~E in Sections~3.1 and 3.2, we believe that M8~E is situated at the edge of the Nebula, formed due to the material being pushed away from the central \hii\ region as  the IF (shown in Fig.~\ref{ysos}) is moving to the south-east. Perusing Figs.~\ref{mean_maps} and \ref{ancillary_data}~(d), we can compare the spatial variation of our observed species with respect to the IF. We find that among all the species discussed in this work, the $^{12}$CO and $^{13}$CO trace the IF the best, while $N_{2}$H$^+$ traces it the poorest. We can also see that HCN, HCO$^{+}$ and HNC emission trace the immediate boundary of the IF both in the north and south, but this trend is not followed by their $^{13}$C isotopes. Although H$^{13}$CN, H$^{13}$CO$^{+}$ and HN$^{13}$C emission trace the southern boundary of the IF, they show no emission toward IF's northern boundary. This observation leads us to an interpretation that the gas toward the south of M8E-IR is denser compared to that to its north.

\subsection{Compression of M8~E and possible triggered star formation?} 


The expansion of the IF is expected to be responsible for  compressing and heating the part of the molecular cloud facing it, which gives rise to the emission of molecular lines as reported in this work. Based on our findings about the IF (bright rimmed structure seen in Fig.~\ref{ancillary_data}~d) and about M8~E environment being denser compared to M8-Main, we suspect that triggered star formation may be at work in M8~E.    

\begin{figure}[htp]
\subfigure{\includegraphics[width=87mm]{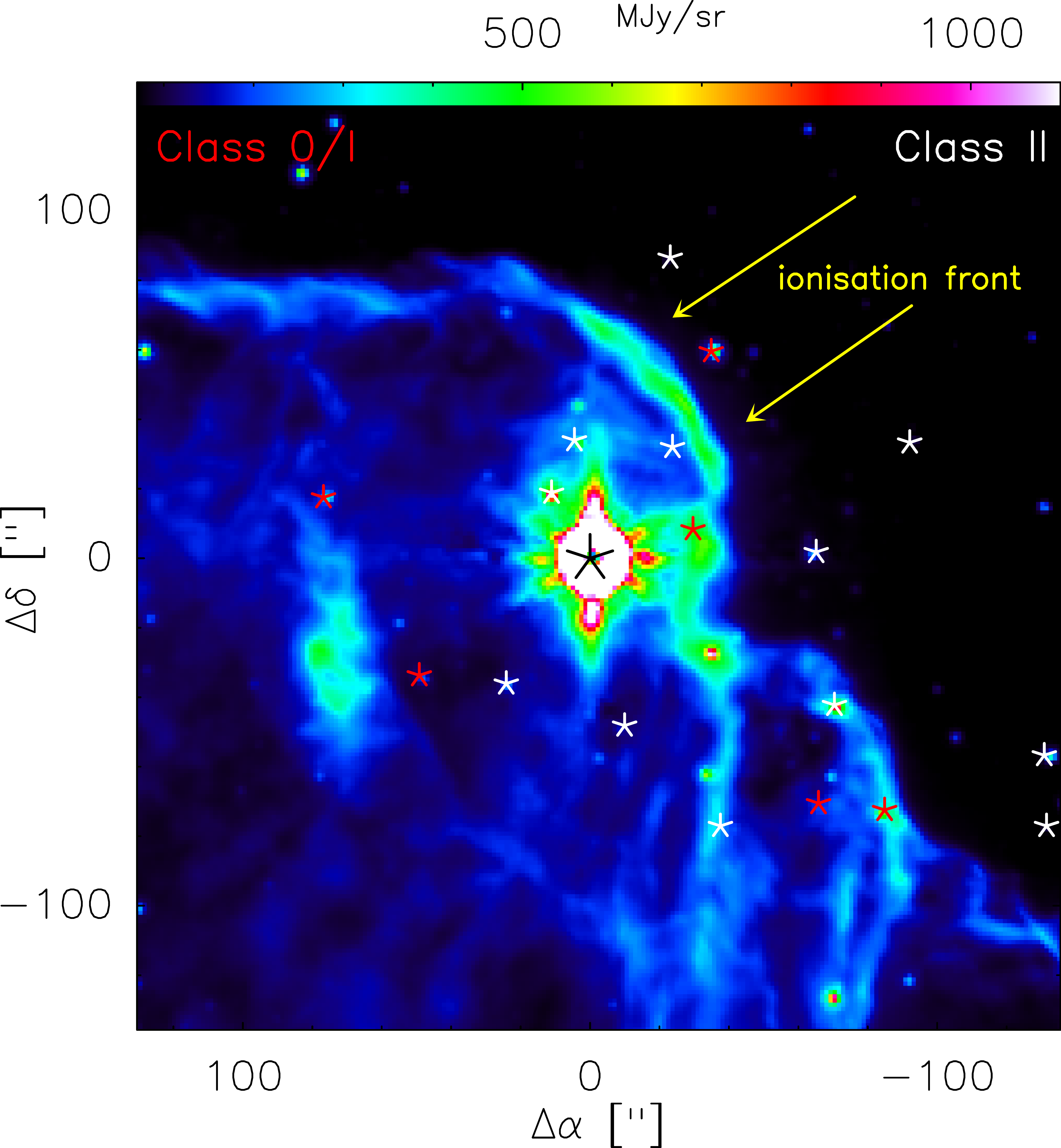}}

\caption{The GLIMPSE~8~$\mu$m PAH emission distribution is shown in color scale along with the Class 0/I (in red) and Class II (in white) population of YSOs in M8~E. The IF, the bright rimmed structure, is pointed out by arrows (in yellow). Position offsets are relative to the position of M8E-IR (marked with an asterisk) given in Section~2. }
  
\label{ysos} 
\end{figure}

In order to investigate the possibility of past triggered star formation in M8~E, we examined the population of young stellar objects in this region. \citet{2010arXiv1005.1148D} catalogued Class 0/I and class I Young Stellar Objects (YSOs)\footnote{\citet{Adams1987} introduced a characterisation for low-mass (solar-type) YSOs, (mainly) based on their IR to submm wavelength spectral energy distributions (SEDs), for which \citet{Lada1991} introduced the nomenclature of Class I--III sources: Class I sources are the youngest, most deeply embedded, whereas Class II sources are classical T Tau stars still surrounded by circumstellar material and, finally, Class III objects that are ``naked'' T Tau stars, that have shed their envelopes. The name for a still earlier phase, Class 0, was coined by \citet{Andre1993}. The  SEDs of these sources show only emission at submm and longer wavelengths.} using  Spitzer/IRAC 4-channel photometry. A total of 27 classical T Tauri stars (CTTs), which are per definition Class II YSOs, have been identified in the region from multi-observatory IR data 
by \citet{2006MNRAS.366..739A, 2007MNRAS.374.1253A}). We list the YSOs that lie close to M8~E in Table~\ref{yso-table}, and Fig.~\ref{ysos} shows their distribution on the GLIMPSE~8~$\mu$m emission image. It can be seen  that the younger YSOs, Class 0/I, predominantly appear to be associated with the dense gas swept up by the IF (east of M8~E), while Class II YSOs (from the older generation) are distributed evenly around both sides of the IF

We note that in their recent study of NGC 6530, conducted in the framework of the Gaia-ESO survey, \citet{Prisinzano2019} also interpreted the distribution of classical T Tauri CTTs relative to the open cluster's O-type stars as evidence for triggered star formation in M8~E, revising the view of \citet{Kalari2015} who, in their earlier work, also studying CTTs in NGC 6530, saw no such evidence.

In order to estimate the velocity of the IF, we assumed that it is moving at a constant speed. The relative distance between the centers of the Class 0/I and Class II distributions of YSOs is about $\sim$ 0.5~pc. We assume a typical upper limit of $\sim$ 2~Myr for the age of Class II YSOs. For the open cluster NGC~6530 \citet{Damiani2019} use Gaia data to find sequential star formation over an age range from 0.5 to 5 Myr; here we adopt 2 Myr. For the age of Class~0/I YSOs we take $\sim$ 0.27~Myr, the average of the ages of Class~0 ($\sim$ 0.1~Myr) and Class~I ($\sim$ 0.44~Myr) YSOs \citep{2009ApJS..181..321E}. With the ages, we determine a lower limit on the speed of the IF $\sim$ 0.26~km~s$^{-1}$, which is in accordance with the maximum predicted IF speed range of $\sim$ 0.5--1~km~s$^{-1}$ in PDR type environments like that of the Orion Bar \citep{1998ApJ...495..853S}.







\section{Conclusions}
We reported 
observations of  diffuse and dense molecular gas tracers toward M8~E in the 3~mm regime using the EMIR receiver of the IRAM~30~m telescope and, for the first time, presented velocity integrated intensity maps of the $J$ = 1 $\to$ 0 transition of a selection of molecules, namely of $^{12}$CO, $^{13}$CO, N$_2$H$^+$, HCN, H$^{13}$CN, HCO$^{+}$, H$^{13}$CO$^{+}$, HNC and HN$^{13}$C in this region.

From ancillary data, analysed in Section~3.1, we found that the 870~$\mu$m submillimeter dust continuum emission probes the dense gas responsible for star formation in M8~E, while the 8~$\mu$m and 22~$\mu$m emission traces the warm gas  In Section~4, using several LTE and non-LTE methods, we determined the warm and cool gas temperatures $\sim$ 80~K and 30~K, respectively and the H$_2$ volume densities in the range of 10$^4$~cm$^{-3}$ to 10$^6$~cm$^{-3}$. Toward M8~E, we employed the molecular cloud thermometer CH$_3$CCH to determine the temperature of the dense gas component. We summarised the differences in the estimated abundances of various species observed toward M8~E and M8-Main, in Section~5, and we believe that star formation is occurring in a denser environment in M8~E compared to M8-Main, placing M8~E at an earlier stage of star formation. Furthermore, the emission from the $J$ = 1 $\to$ 0 transition of N$_2$H$^+$ and HN$^{13}$C is spatially most concentrated and is consistent with dense and cool gas (as probed by CH$_3$CCH).


From the geometry of the region and the distribution of the YSOs in M8~E, we find that large quantity of gas is being compressed and swept away by the IF of the expanding central \hii\ region of M8. Based on the ages of different classes of the YSO population found in M8~E, we estimate that the IF is moving at least at a speed of $\sim$ 0.26~km~s$^{-1}$. Hence, we conclude that an earlier generation of stars in the cluster NGC6530 is responsible for the triggered star formation we observe in M8~E.

\section{Outlook}
Compared to the power house O stars that excite the bulk of M8 (see Section~\ref{intro}), the young stellar objects embedded in M8~E, with spectral types equivalent to B0 and B2, together with a few sources of still lower luminosity, clearly have a much smaller impact on the region, now and in the future.  Nevertheless, as shown here, studies of M8 and its environs, with M8~E the so far best studied example, allow interesting studies of the outcome of recent (M8-Main) and ongoing star formation that is triggered by the former.

The purpose of this paper was to map out the extent of the molecular material in M8~E and to derive its basic properties by analyzing data from ground state transitions of common molecules. Higher-$J$ lines of these and other molecules measured with the APEX telescope will allow a comprehensive characterization of the energetics, density and chemistry of the molecular gas, including the powerful outflow. In particular, the source's chemical richness  will be explored by data obtained in deep integrations toward M8E-IR, which for the 3 mm band are already in hand. 

Much of the gaseous content of the wider M8 region remains unexplored, along with  embedded YSOs that it may harbour, and warrants further exploration: Looking at the 450~$\mu$m continuum image (Fig.~\ref{m8_overview}), one clearly notices many other dust condensations, whose star forming activity has not yet been investigated, for example those in the striking $\sim$ 5~pc long curved string that has M8~E at its eastern end and the $\sim 1$~pc long north-south ridge abutting NGC 6530. Future systematic studies of the M8 region might discover objects comparable to M8 E, if less luminous, and in yet earlier stages of star formation and hold the promise to find more evidence for triggered star formation.

\begin{acknowledgements}
M. Tiwari was supported for this research by the International Max Planck Research School (IMPRS) for Astronomy and Astrophysics at the Universities of Bonn and Cologne. We thank Nina Brinkmann for helpful discussions. 
\end{acknowledgements}



\clearpage

\bibliography{references1}
\bibliographystyle{aa}

\clearpage
\begin{appendix}

\section{Spatial distribution of the 26~km~s$^{-1}$ molecular cloud}
\begin{figure}[htp]
\centering
\subfigure{\includegraphics[width=80mm]{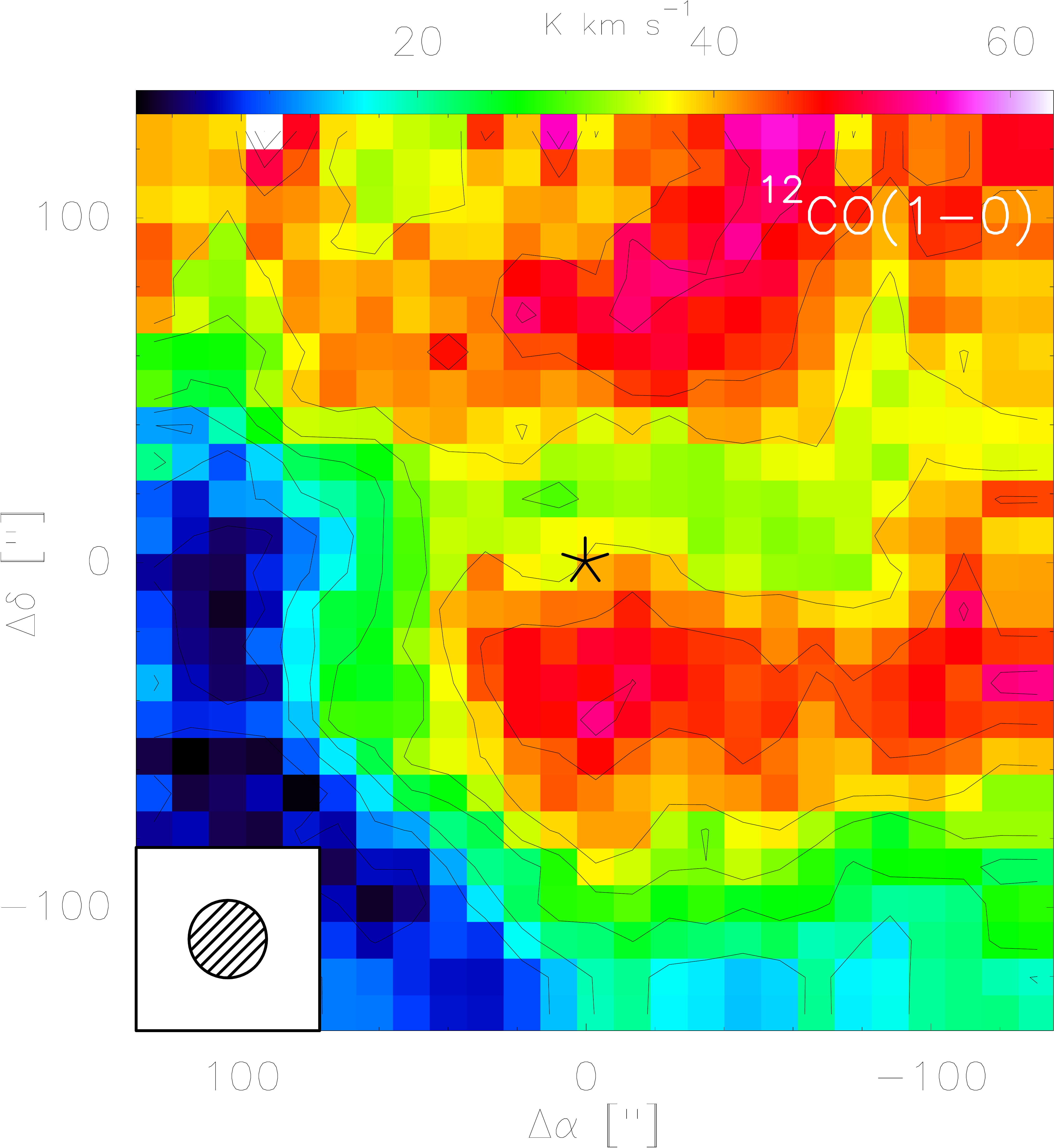}}
\subfigure{\includegraphics[width=80mm]{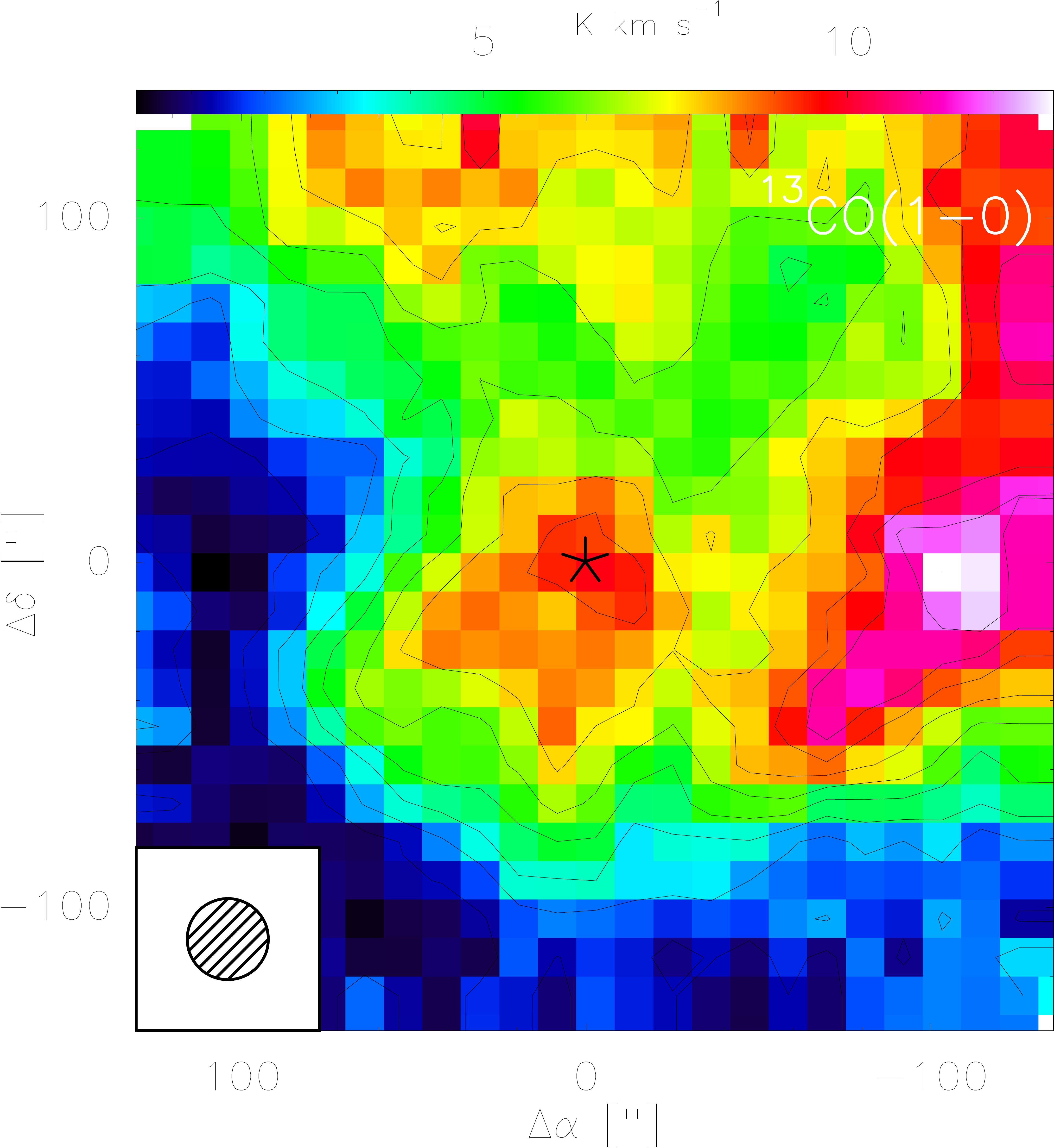}}

\caption{Intensity color maps of the $J$ = $1 \to 0$ transition of $^{12}$CO and $^{13}$CO integrated within the velocity range of 22--30~km~s$^{-1}$. Position offsets are relative to the position of M8E-IR (marked with an asterisk) given in Section~2. }
  
\label{v26} 
\end{figure}

Velocity integrated intensity maps of $J$ = 1 $\to$ 0 transitions of $^{12}$CO and $^{13}$CO for the molecular cloud with $\varv_{\rm LSR} = 26~$km~s$^{-1}$ are shown in Fig.~\ref{v26}.

Similar to the main molecular cloud of M8~E (at $\varv_{\rm LSR} = 11$~km~s$^{-1}$), which  is associated with M8-Main, the emission from the $+26$~km~s$^{-1}$ molecular cloud has a local maximum toward M8E-IR. But unlike the main molecular cloud whose emission is abruptly stops south-west of M8E-IR due to M8 Main's influence, 
the emission distribution of this cloud, which is located along the line of sight to M8, has a much more widespread distribution.

Using the Revised Kinematic Distance Calculator accessible on the website of the Bar and Spiral Structure Legacy (BeSSeL) Survey\footnote{\url{http://bessel.vlbi-astrometry.org/revised_kd_2014}}
that uses values for the Galactic parameters published by \citet{Reid2014}, for $\varv_{\rm LSR} = 26$~km~s$^{-1}$ we calculate near and far heliocentric kinematic distances of $4.08^{+0.55}_{-0.73}$ kpc and $12.05^{+0.58}_{-0.45}$ kpc, respectively, while for  $\varv_{\rm LSR} = 11$~km~s$^{-1}$ we find a (near) distance of $2.15^{+1.07}_{-1.62}$ kpc, which is consistent with the parallax distance for M8 (1.35 kpc).

\section{YSO population in M8~E}

\begin{table}[h]
\tiny
\centering
\begin{threeparttable}
\caption{Identified YSOs in M8~E as reported in \cite{2010arXiv1005.1148D}.}
\renewcommand{\arraystretch}{1.4}
\begin{tabular}{c c c}
\hline\hline
  Right Ascension (J2000) & Declination (J200) & Object\tnote{a} \\
  \hline
\multicolumn{3}{c}{Class~0/I}  \\ 
 \hline
 18:04:47.09 & -24:27:55.4 & 58\\
 18:04:48.42 & -24:27:53.8 & 59\\
 18:04:50.37 & -24:14:25.5 & 60\\
 18:04:50.62 & -24:25:42.2 & 61\\
 18:04:56.77 & -24:27:16.4 & 63\\
 18:04:58.82 & -24:26:24.1 & 64\\
 \hline
 \multicolumn{3}{c}{Class~II}  \\ 
 \hline
 18:04:43.53 & -24:27:38.7 & 152\\
 18:04:43.65 & -24:27:59.1 & 153\\
 18:04:46.41 & -24:26:08.0 & 156\\
 18:04:48.05 & -24:27:24.6 & 157\\
 18:04:48.56 & -24:26:40.7 & 158\\
 18:04:50.23 & -24:27:59.4 & 159\\
 18:04:51.53 & -24:24:17.4 & 160\\
 18:04:51.57 & -24:26:10.9 & 161\\
 18:04:52.64 & -24:27:30.8 & 163\\
 18:04:53.48 & -24:26:08.7 & 164\\
 \hline
\end{tabular}

\begin{tablenotes}
\item[a]\small Refers to the YSO number as mentioned in the catalogue reported by \cite{2010arXiv1005.1148D}.
\end{tablenotes}

 \label{yso-table}
 \end{threeparttable}
\end{table}

\end{appendix}

\end{document}